\documentclass[prd,10pt,nofootinbib,twocolumn,superscriptaddress,preprintnumbers,aps]{revtex4-1}

\usepackage{xcolor,colortbl}
\usepackage{bbold}

\definecolor{Gray}{gray}{0.95}
\definecolor{RGray}{gray}{0.85}
\definecolor{CGray}{gray}{0.92}
\newcolumntype{a}{>{\columncolor{Gray}}c}
\newcolumntype{b}{>{\columncolor{white}}c}

\newcommand{\RDDs}{$R(D^{(*)})$}
\newcommand{\RDs}{$R(D^*)$}

\usepackage{amssymb}
\usepackage{amsmath}
\usepackage{graphicx}
\usepackage{verbatim}
\usepackage{amsfonts}

\usepackage[utf8]{inputenc}

\newcommand{\be}{\begin{equation}}
\newcommand{\ee}{\end{equation}}
\newcommand{\beqn}{\begin{eqnarray}}

\newcommand{\bat}{\begin{array}{cc}}
\newcommand{\ea}{\end{array}}
\newcommand{\batt}{\begin{array}{ccc}}
\newcommand{\eeqn}{\end{eqnarray}}
\newcommand{\cL}{{\cal L}}
\newcommand{\BtoDDs}{B\to D^{(*)}\tau \nu}

\usepackage{color}
\definecolor{color2}{rgb}{0,0.5,0.5}
\definecolor{color3}{rgb}{0.7,.2,.7}
\definecolor{color4}{rgb}{0.9,.4,.4}

\allowdisplaybreaks

\begin{document}

\vspace*{-30mm}

\title{Scalar contributions to $b\to c (u) \tau \nu$ transitions}
\author{Alejandro Celis}
\affiliation{Ludwig-Maximilians-Universit\"at M\"unchen,
   Fakult\"at f\"ur Physik,\\
   Arnold Sommerfeld Center for Theoretical Physics,
   80333 M\"unchen, Germany}
\author{Martin Jung}
\affiliation{TUM Institute for Advanced Study, Lichtenbergstr.~2a, D-85747 Garching, Germany}
\affiliation{Excellence Cluster Universe, Technische Universit\"at M\"unchen, Boltzmannstr. 2, D-85748 Garching, Germany}
\author{Xin-Qiang Li}
\affiliation{Institute of Particle Physics and Key Laboratory of Quark and Lepton Physics~(MOE),\\
Central China Normal University, Wuhan, Hubei 430079, P.~R.~China}
\author{Antonio Pich}
\affiliation{IFIC, Universitat de Val\`encia -- CSIC, Apt. Correus 22085, E-46071 Val\`encia, Spain}

\vspace*{1cm}

\begin{abstract}\noindent
We perform a comprehensive analysis of scalar contributions in $b \to c \tau \nu$ transitions including the latest measurements of $R(D^{(*)})$, the $q^2$ differential distributions in $B \to D^{(*)} \tau \nu$, the $\tau$ polarization asymmetry for $B \to D^{*} \tau \nu$, and the bound derived from the total width of the $B_c$ meson. We find that scalar contributions with the simultaneous presence of both left- and right-handed couplings to quarks can explain the available data, specifically $R(D^{(*)})$ together with the measured differential distributions. However, the constraints from the total $B_c$ width present a slight tension with the current data on $B \to D^{*}\tau \nu$ in this scenario, preferring smaller values for $R(D^*)$. We discuss possibilities to disentangle scalar new physics from other new-physics scenarios like the presence of only a left-handed vector current, via additional observables in $\BtoDDs$ decays or additional decay modes like the baryonic $\Lambda_b \to
\Lambda_c \tau \nu$ and the inclusive $B \to X_c \tau \nu$ decays. We also analyze scalar contributions in $b \to u \tau \nu$ transitions, including the latest measurements of $B \to \tau \nu$, providing predictions for $\Lambda_b \to p \tau \nu$ and $B \to \pi \tau \nu$ decays. The potential complementarity between the $b \to u$ and $b \to c$ sectors is finally investigated once assumptions about the flavour structure of the underlying theory are made.
\end{abstract}

\preprint{LMU-ASC 57/15}
\preprint{FLAVOUR(267104)-ERC-112}
\preprint{IFIC/16-92}

\maketitle


\section{Introduction} \label{sec:intro}

The first run of the Large Hadron Collider (LHC) has completed experimental evidence for the Standard Model (SM) of electroweak (EW)
interactions by  discovering a scalar boson with properties consistent with a SM Higgs doublet~\cite{Aad:2012tfa,*Chatrchyan:2012ufa}.
The absence of clear signals beyond the SM in both collider and flavour analyses seems to suggest that the scale of new physics (NP)
is much higher than the EW scale. However, relatively light weakly-coupled particles could have been missed by current searches, given the limited sensitivity of the LHC to such states. In particular, additional light scalar bosons, predicted in many extensions of
the SM, are in general still allowed.

\begin{figure}[thb]
\centering
\includegraphics[width=8cm,height=8cm]{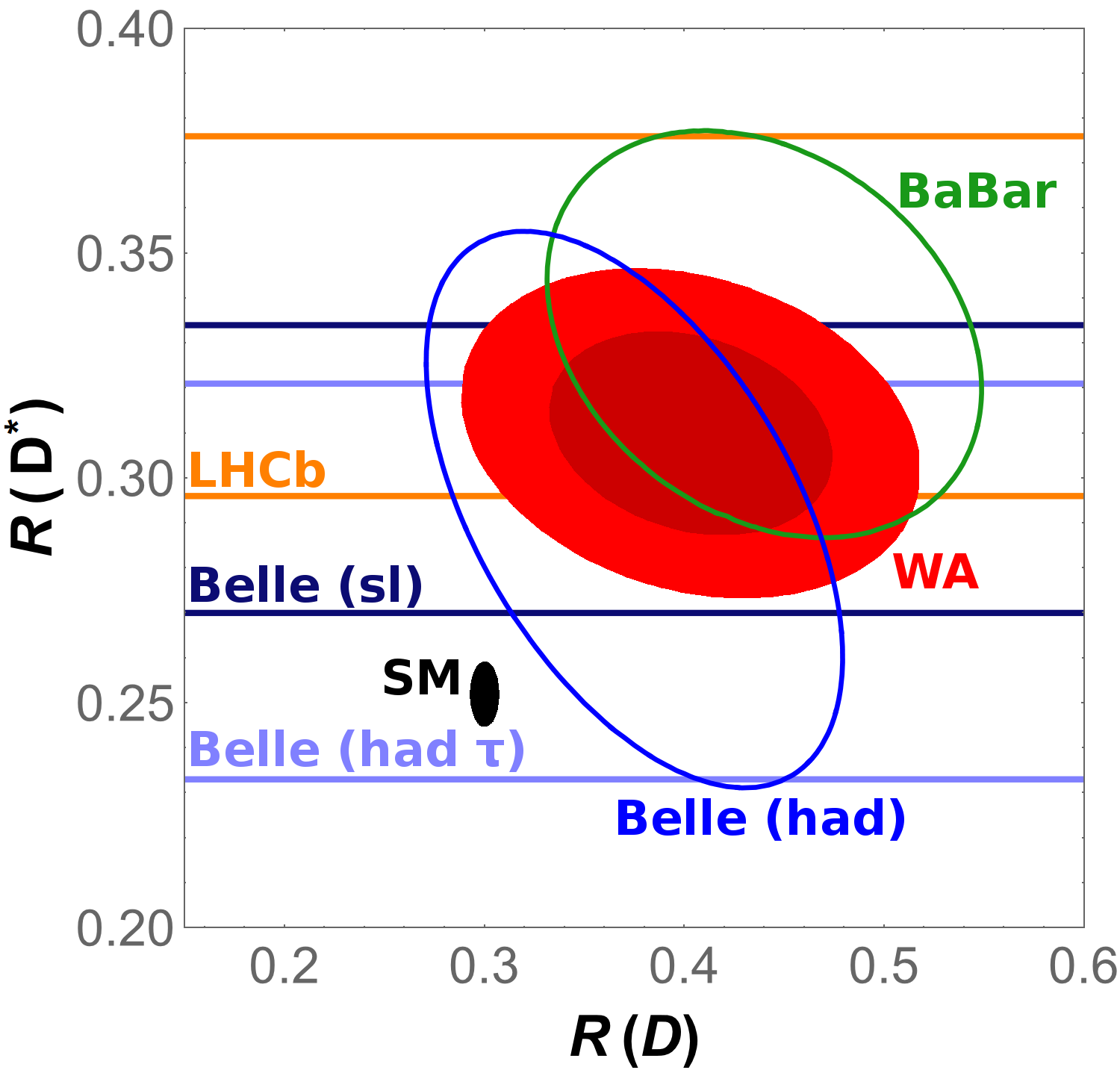}
\caption{ \it \small Average of \RDDs{} measurements, displayed as red filled ellipses ($68\%$~CL and $95\%$~CL). The SM prediction is shown as a black ellipse ($95\%$~CL), and the individual measurements as continuous contours ($68\%$~CL): Belle (blue ellipse and horizontal bands), BaBar (green ellipse), and LHCb (horizontal orange band).
\label{fig:expdata}
}
\end{figure}

In this work we are interested in the possibility of sizable scalar couplings in $b\to c (u) \tau \nu$ transitions, as induced for instance by a charged-scalar boson with a mass around the EW  scale~\cite{Hou:1992sy,*Tanaka:1994ay,*Kiers:1997zt,*Nierste:2008qe,*Kamenik:2008tj,*Tanaka:2010se,Jung:2010ik}.
In 2012 the BaBar collaboration observed an excess in $B\to D^{(*)} \tau \nu$ decays with respect to the SM predictions, hinting at a
violation of lepton-flavour universality at the $30\%$ level~\cite{Lees:2012xj}. The measured observables are the ratios
\be\label{eq:RDdef}
R(D^{(*)}) = \frac{\mathrm{Br}(\BtoDDs)}{\mathrm{Br}(B \to D^{(*)} \ell \nu)}\,,
\ee
with $\ell=e$ or $\mu$, in which many sources of experimental as well as theoretical uncertainties cancel. These deviations cannot be
accommodated by a charged-scalar contribution in the type-II two-Higgs-doublet model (2HDM)~\cite{Lees:2012xj}, motivating the discussion of more general extended scalar sectors as well as different NP interpretations~\cite{Fajfer:2012jt,Datta:2012qk,*Tanaka:2012nw,*He:2012zp,*Sakaki:2013bfa,*Biancofiore:2013ki,*Dutta:2013qaa,*Duraisamy:2014sna,*Hagiwara:2014tsa,*Bhattacharya:2015ida,*Alonso:2016gym,*Dumont:2016xpj,*Das:2016vkr,*Ivanov:2016qtw,*Sahoo:2016pet,*Becirevic:2016yqi,*Ko:2012sv,*Abada:2013aba,*Dorsner:2013tla,*Enomoto:2015wbn,*Hati:2015awg,*Ligeti:2016npd,*Bardhan:2016uhr,*Bhattacharya:2016mcc,*Hiller:2016kry,*Bauer:2015knc,*Feruglio:2016gvd,*Barbieri:2015yvd,*Deshpand:2016cpw,*Barbieri:2016las,*Boubaa:2016mgn,*Becirevic:2016hea,*Alok:2016qyh,*Nandi:2016wlp,*Wang:2016ggf,*Dutta:2016eml,*Bhattacharya,*Hue:2016nya,Bhattacharya:2014wla,Greljo:2015mma,Calibbi:2015kma,Boucenna:2016wpr,Boucenna:2016qad,Celis:2012dk,Crivellin:2013wna,Sakaki:2014sea,Alonso:2015sja,Freytsis:2015qca,Crivellin:2015hha,Cline:2015lqp,Li:2016vvp}.
Recently, the LHCb collaboration announced a measurement of \RDs{}~\cite{Aaij:2015yra}, and the Belle collaboration published
several analyses with different decays on the tagging side as well as different $\tau$-decay final states: an update of their analysis
of \RDDs{} with hadronic tagging and leptonic $\tau$ decay~\cite{Huschle:2015rga}, an analysis of $R(D^*)$ with semileptonic tagging and leptonic $\tau$ decay~\cite{Sato:2016svk}, and the first analysis of \RDs{} with hadronic tagging and different hadronic
$\tau$-decay final states~\cite{Hirose:2016wfn} which importantly includes for the first time a measurement of the $\tau$ polarization
in this mode, albeit with rather limited precision. All available measurements~\cite{Amhis:2016xyh} are very consistent and, while
all recent analyses are individually compatible with the SM predictions at $\sim 95\%$~CL, they yield the average
\be \label{eq::RDDsavg}
R(D) = 0.403\pm0.047\,,\quad
R(D^*) = 0.310\pm0.015\,,
\ee
with a correlation of $-23\%$, as displayed in Fig.~\ref{fig:expdata}.
This implies a deviation from the SM predictions of about $4\sigma$. Furthermore, the shapes of the differential distributions
$d\Gamma(\BtoDDs)/dq^2$ have been made available by Belle~\cite{Huschle:2015rga} and BaBar~\cite{Lees:2013uzd}, yielding additional information to distinguish NP from the SM as well as different NP models from each other. We also include information from the inclusive decay $b\to X\tau\nu$, measured at LEP~\cite{Abbaneo:2001bv}. Finally, the total width of the $B_c$ meson can help to exclude fine-tuned solutions with very large NP contributions~\cite{Li:2016vvp,Alonso:2016oyd}. The possibility of scalar contributions in $b\to u \tau\nu$ transitions is also analyzed, paying in particular attention to the potential complementarity between the $b\to c$ and $b\to u$ sectors.

Our paper is organized as follows: In Sec.~\ref{sec:frame} we present the theoretical framework used in this work. The physical observables considered in our analysis are summarized in Sec.~\ref{sec:exp}. In Sec.~\ref{sec:disc} we discuss the phenomenological implications of current data, before concluding in Sec.~\ref{sec:conc}. Hadronic input parameters and the statistical treatment are discussed in Appendix~\ref{sec:inputs}. Details on the $b\to c\tau\nu$ observables entering our analysis, like the $q^2$ distributions for $\BtoDDs$ decays, are collected in Appendix~\ref{app:diff}. Details of the fit are provided in Appendix~\ref{fitdetl}.

\section{Framework} \label{sec:frame}

The study of NP contributions to charged-current semileptonic processes can in principle be carried out in a model-independent manner.
We discuss here the subset of operators induced \emph{e.g.} by charged scalars which naturally lead to observable effects in $b\to
c (u) \tau\nu$ transitions, while $b\to c (u) \ell\nu$ remain unaffected. Note that in general this is not true for contributions from
left- or right-handed vector currents. Specifically, right-handed vector currents are explicitly lepton-flavour-universal in all models
with SM particle content and gauge symmetry at the EW scale, if the EW symmetry is linearly realized, up to contributions of order
$v^4/\Lambda^4$, where $v$ denotes the EW vacuum expectation value and $\Lambda$ the scale of additional NP
particles~\cite{Cata:2015lta,Alonso:2015sja,Cirigliano:2009wk}.
While this universality can be broken if the EW symmetry is non-linearly realized~\cite{Cata:2015lta}, right-handed vector-current
contributions are  generally strongly constrained by semileptonic $B$ decays into light lepton modes. When comparing with NP scenarios
with vector couplings, we therefore assume vanishing right-handed couplings.

The low-energy effective Lagrangian describing scalar-mediated charged-current semileptonic transitions is given by
\be \label{eq:Lag}
\cL_{\mbox{\scriptsize{eff}}} =-\frac{4 G_F  V_{q_uq_d}  }{\sqrt{2}} \left[ \bar q_u (  g_{L}^{q_uq_d\ell} \mathcal{P}_L  + g_{R}^{q_uq_d\ell}  \mathcal{P}_R   ) q_d   \right] [\bar \ell  \mathcal{P}_L  \nu_{\ell}] ,
\ee
where we neglect neutrino-mass-related terms with right-handed neutrinos, $V$ represents the Cabibbo-Kobayashi-Maskawa (CKM)
matrix~\cite{Cabibbo:1963yz,*Kobayashi:1973fv}, and $\mathcal{P}_{L,R}=(1\mp\gamma_5)/2$ are the usual chiral projectors. The Wilson
coefficients $g_{L,R}^{q_uq_d\ell}$ are complex parameters which encode details of the theory at high energies. Note that the explicit
appearance of the CKM matrix does not imply any assumption about the flavour structure of the underlying theory at this stage, but is
merely a choice of normalization of the Wilson coefficients. They are in full generality independent for every possible flavour
combination $q_u=(u,c,t)$, $q_d=(d,s,b)$, $\ell=(e,\mu,\tau)$, yielding 54 couplings. However, Eq.~\eqref{eq:Lag} effectively already
assumes a colour-neutral scalar exchange, since generally a coloured scalar like a leptoquark would induce tensor couplings as
well~\cite{Dorsner:2016wpm}. Therefore, without (further) loss of generality, we can assume the couplings to obey
\be
g_{L,R}^{q_uq_d\ell}=g_{L,R}^{q_uq_d}g_{L}^\ell \,,
\ee
thereby reducing the number of independent parameters to 21: two general matrices in quark-flavour space $g_{L,R}^{q_uq_d}$ and three couplings $g_L^\ell$. Since we assume that the NP effects are negligible for the light lepton modes, we set $g_L^{e,\mu}=0$. Considering $b \to c (u)$ transitions restricts the quark-sector parameters in our analysis to $g_{L,R}^{c(u)b}$, \emph{i.e.} 4 complex couplings. This effective Lagrangian allows for a model-independent discussion of scalar contributions in $b\to c (u) \tau\nu$ transitions,
which comprises the objective of our analysis. This general scenario will be dubbed S1 in the following. A particular realization of this framework is provided by the type-III 2HDM, see \emph{e.g.} Refs.~\cite{Branco:2011iw,Crivellin:2012ye,Crivellin:2013wna,Cline:2015lqp} for recent discussions.\footnote{Note that the
interpretation of the $R(D^{(*)})$ anomalies in terms of a 2HDM is severely constrained by the LHC searches for additional scalars in the $\tau^+\tau^-$ channel~\cite{Cline:2015lqp,Faroughy:2016osc}.}

If we want to relate processes involving different flavour transitions, we need to make assumptions about the flavour structure of the
underlying theory. In order to study the potential complementarity between $b \to c$ and $b \to u$ probes of scalar contributions, we will consider as a benchmark the \emph{universality} relations\footnote{\emph{Universality} refers here to the relation between the two sets of Yukawa matrices occurring in 2HDMs, \emph{i.e.} $Y^{u,d}_1\sim Y^{u,d}_2$, leading to scalar couplings to fermions proportional to the fermion masses.}
\be \label{eq:rel}
\frac{g_{L}^{cb}}{g_{L}^{ub}} = \frac{m_{c}}{m_{u}} \,,
\qquad\qquad
\frac{g_{R}^{cb}}{g_{R}^{ub}} = 1 \,,
\ee
which are realized \emph{e.g.} in 2HDMs with natural flavour conservation (NFC)~\cite{Paschos:1976ay,Glashow:1976nt}, but also \emph{e.g.} in the aligned 2HDM~\cite{Pich:2009sp,Jung:2010ik}. This scenario will be labelled S2 in the following. In our analysis we will consider both scenarios S1 and S2 with complex as well as real parameters.

\section{Observables} \label{sec:exp}

\begin{table*}[bht]
\centering{
\caption{\label{TAB:EXP} \it \small Predictions within the SM for the various leptonic and semileptonic decays considered in this work, together with their corresponding experimental values. ${}^\dagger$The correlation between $R(D)$ and $R(D^*)$ is $-23\%$.} \vspace{0.2cm}
\doublerulesep 0.8pt \tabcolsep 0.02in
\begin{tabular}{cccc}\hline\hline   \rowcolor{RGray}
Observable & SM prediction & Exp. value &  Reference \\ \hline
$ R(D)$                         &
$0.301 \pm 0.003$     &    $0.403\pm0.047$           &
\cite{Bozek:2010xy,Lees:2012xj,Huschle:2015rga,Amhis:2016xyh}
 \\    \rowcolor{Gray}
$R(D^*)$                        &
$0.252\pm0.001\pm0.003$   &  $0.310\pm0.015^\dagger$          &
\cite{Bozek:2010xy,Lees:2012xj,Aaij:2015yra,Huschle:2015rga,Hirose:2016wfn,Amhis:2016xyh} \\
$A_\lambda(D^*)$ & $0.502\pm0.005\pm0.017$ & $0.38\pm0.55$ &
\cite{Hirose:2016wfn}\\\rowcolor{Gray}
$R(X_c)$ & $0.222\pm0.000\pm0.007$ & $0.220\pm0.022$ &
\cite{Olive:2016xmw}\\
$R(\tau)$  &  $0.48\pm0.04$  & $0.72\pm0.13$ & \cite{Amhis:2016xyh,Lees:2012ju,*Aubert:2009wt,*Adachi:2012mm,*Kronenbitter:2015kls} \\\rowcolor{Gray}
$R(\pi)$  &
$0.594^{+0.017}_{-0.015}$  &  $1.03\pm 0.49\; (\leq 2.0)$       & \cite{Hamer:2015jsa,Amhis:2016xyh} \\
\hline\hline
\end{tabular}
}
\end{table*}

The low-energy flavour processes considered in this work are summarized in Table~\ref{TAB:EXP}, together with their corresponding SM predictions
and the current experimental values. In addition to the changes discussed in the introduction, these values include new measurements of
the branching ratio for $B \to \tau \nu$ by BaBar and Belle~\cite{Kronenbitter:2015kls}, and a very recent upper limit from Belle on
$R(\pi) = \mathrm{Br}(B\to\pi\tau\nu)/\mathrm{Br}(B\to\pi\ell\nu)$~\cite{Hamer:2015jsa}. Explicit formulae for all these observables
taking into account the scalar contribution have been provided in Refs.~\cite{Jung:2010ik,Celis:2012dk};\footnote{Our definition for
the $\tau$ polarization asymmetry $A_\lambda(D^*)$ differs by a global sign from the one used by the Belle collaboration in
Ref.~\cite{Hirose:2016wfn}.} the necessary adaption of the expression for $B\to\pi\tau\nu$ is discussed in Appendix~\ref{sec:inputs}.
Note that instead of using the branching ratio of $B\to\tau\nu$ directly, we normalize it to that of
$B\to\pi\ell\nu$~\cite{Chen:2006nua,Khodjamirian:2011ub,Fajfer:2012jt}:
\begin{equation}\label{eq::Rtau}
R(\tau)\equiv \frac{\mathrm{Br}(B\to\tau\nu)}{\mathrm{Br}(B\to\pi\ell\nu)}\,.
\end{equation}
While this normalization does not yield any advantage experimentally, it yields the cancellation of $|V_{ub}|$ which is very helpful,
given the discrepancy between the inclusive and exclusive determinations of this
quantity at present, see the article by Kowalewski and Mannel in Ref.~\cite{Olive:2016xmw} for a review.

The $q^2$ distributions for $B\to D^{(*)} \tau\nu $ are given in Appendix~\ref{app:diff}, where also their treatment within the present
analysis is described. Importantly, we leave the normalization of each of these distributions free in the fit, thereby decoupling the
information from the shapes of the distributions from that of the measurements of \RDDs{}, which are already included in the averages
in Eq.~\eqref{eq::RDDsavg}. We introduce binned quantities $R(D^{(*)},i)$ in analogy with Eq.~\eqref{eq:RDdef} as
\begin{equation}\label{eq::RDDsbinned}
R(D^{(*)},i) \equiv \frac{\int_{{\rm bin}\, i}dq^2\;\frac{d\Gamma(\BtoDDs)}{dq^2}}{\int_{{\rm bin}\, i}dq^2\;\frac{d\Gamma(B\to D^{(*)}\ell
\nu)}{dq^2}}\,.
\end{equation}
The binning is given with the experimental data in Tables~\ref{tab:distBelle} and~\ref{tab:distBaBar} in the appendix.

The experimental values for the differential distributions and \RDDs{} depend on the size of the potential NP contribution, since
the latter affects the kinematics of the decay distribution~\cite{Lees:2013uzd}. We will comment on this issue when performing the fits in the next section.

Note that the measured values of \RDDs{} oversaturate the SM prediction for the inclusive $B \to X_c \tau \nu$ decay rate when
including an estimate for the decays to other excited charm-meson states, implying that the tension in \RDDs{} with the SM predictions
is independent of the $B\to D^{(*)}$ form-factor determination~\cite{Ligeti:2014kia,Freytsis:2015qca}. Furthermore, the $\BtoDDs$ modes
already saturate the inclusive branching ratio ${\mathrm{Br}}(b\to X_c\tau\nu)$ that can be estimated from the LEP measurement of
$b$-hadron decays to final states with a $\tau$ lepton. A confirmation of the latter result with higher precision would indicate that
the actual value for \RDDs{} is smaller than the present average, closer to the Belle central value. Below we discuss the inclusive
measurement without relying on estimates for the decays to excited charm-meson states. We calculate $R(X_c) = \mathrm{Br}(B \to X_c
\tau \nu)/\mathrm{Br}(B \to X_c \ell \nu)$ consistently at next-to-leading order (NLO), which results in a qualitative difference for
the non-SM part compared to the leading-order (LO) result. Details of the calculation are deferred to Appendix~\ref{app:diff}.

The limit from the total width of the $B_c$ meson is obtained as follows: we consider only the modification due to the decay
$B_c\to\tau\nu$, which is calculable once the decay constant is known. To this end we add an estimate for those $B_c$ decays which are
modified negligibly by scalar NP. Apart from the fact that NP models with new scalar interactions typically yield charged-scalar
interactions that are at least roughly proportional to the fermionic mass, this is justified by the very successful SM predictions of
leptonic $\pi$, $K$ and $D$ decays: large corrections to the light-lepton or first-family quark couplings would be visible in these
modes. Given that they make up over $85\%$ of the successfully predicted total width~\cite{Beneke:1996xe}, we consider an upper limit
$\mathrm{Br}(B_c\to\tau\nu)\leq 40\%$, which is still extremely conservative and thereby accounts also for sizable theory
uncertainties in this estimate;\footnote{We observe that our results are not affected in a significant manner by using instead the
slightly stronger limit $\mathrm{Br}(B_c\to\tau\nu)\lesssim 30\%$ used in Ref.~\cite{Alonso:2016oyd}.} note that its SM value is about
$2\%$.

For the baryonic decays $\Lambda_b \to p \ell \nu$ and $\Lambda_b \to \Lambda_c \ell \nu$ we follow Refs.~\cite{Detmold:2015aaa,Shivashankara:2015cta,Dutta:2015ueb,Li:2016pdv}.

Further useful measurements of $b\to c (u) \tau\nu$ transitions include the branching ratios of $B_s\to D_s^{(*)}\tau\nu$, $B_c\to J/\psi\tau\nu$, $B_s\to K^{(*)}\tau\nu$, and $B\to D^{**}\tau\nu$ decays; the hadronic uncertainties for these modes are, however, not yet on the same level as for the observables discussed in this
work.

\section{Discussion} \label{sec:disc}

We now discuss the implications of current flavour data for the couplings in the effective Lagrangian in Eq.~\eqref{eq:Lag}, first model-independently (scenario S1) and then imposing the universality relations in Eq.~\eqref{eq:rel} (scenario S2). We focus on the new elements in our analysis, \emph{i.e.} the influence of the new data for \RDDs, the differential distributions in $\BtoDDs$,
the inclusive mode $b\to X\tau\nu$, the total width of the $B_c$ meson, and the interplay with $b\to u \tau\nu$ transitions. For the scenarios that remain viable we give predictions for selected additional observables that could be measured in the future.

\subsection{Model-independent analysis -- S1}  \label{mia}

\subsubsection{$\mathbf{\boldsymbol{b}\to\boldsymbol{c \tau \nu}}$}

Given the discrepancy of \RDDs{} with respect to the SM predictions we start by analyzing the possibility of accommodating $B \to D^{(*)} \tau \nu$ data by a scalar contribution. Without assumptions on the flavour structure, only observables corresponding to $b\to c\tau\nu$ transitions can be included model-independently. These are the available observables from $\BtoDDs$, $R(X_c)$, and the total width of the $B_c$ meson. Note that $B\to D\tau\nu$ and $B\to D^*\tau\nu$ depend only on the parameter combinations
\be\label{eq::deltas}
\delta_{cb}^{\ell}
\equiv \frac{       (    g_L^{cb\ell}    + g_R^{cb\ell}  )(m_B- m_D)^2 }{  m_{\ell}\,   (  \bar m_b - \bar m_c )} ,\,\,
\Delta_{cb}^{\ell} \equiv \frac{     (   g_{L}^{cb\ell}    - g_{R}^{cb\ell} ) m_B^2  }{   m_{\ell}\,    ( \bar m_b + \bar m_c      ) } ,
\ee
respectively, which we consequently choose to display the corresponding constraints. This implies that any value of $R(D)$ and $R(D^*)$ can at first be trivially explained in this scenario. However, the remaining observables give independent constraints, potentially allowing to rule out scalar NP as an explanation of the observed anomaly.

\begin{figure*}[hbt]
\centering
\includegraphics[width=7.5cm,height=7.5cm]{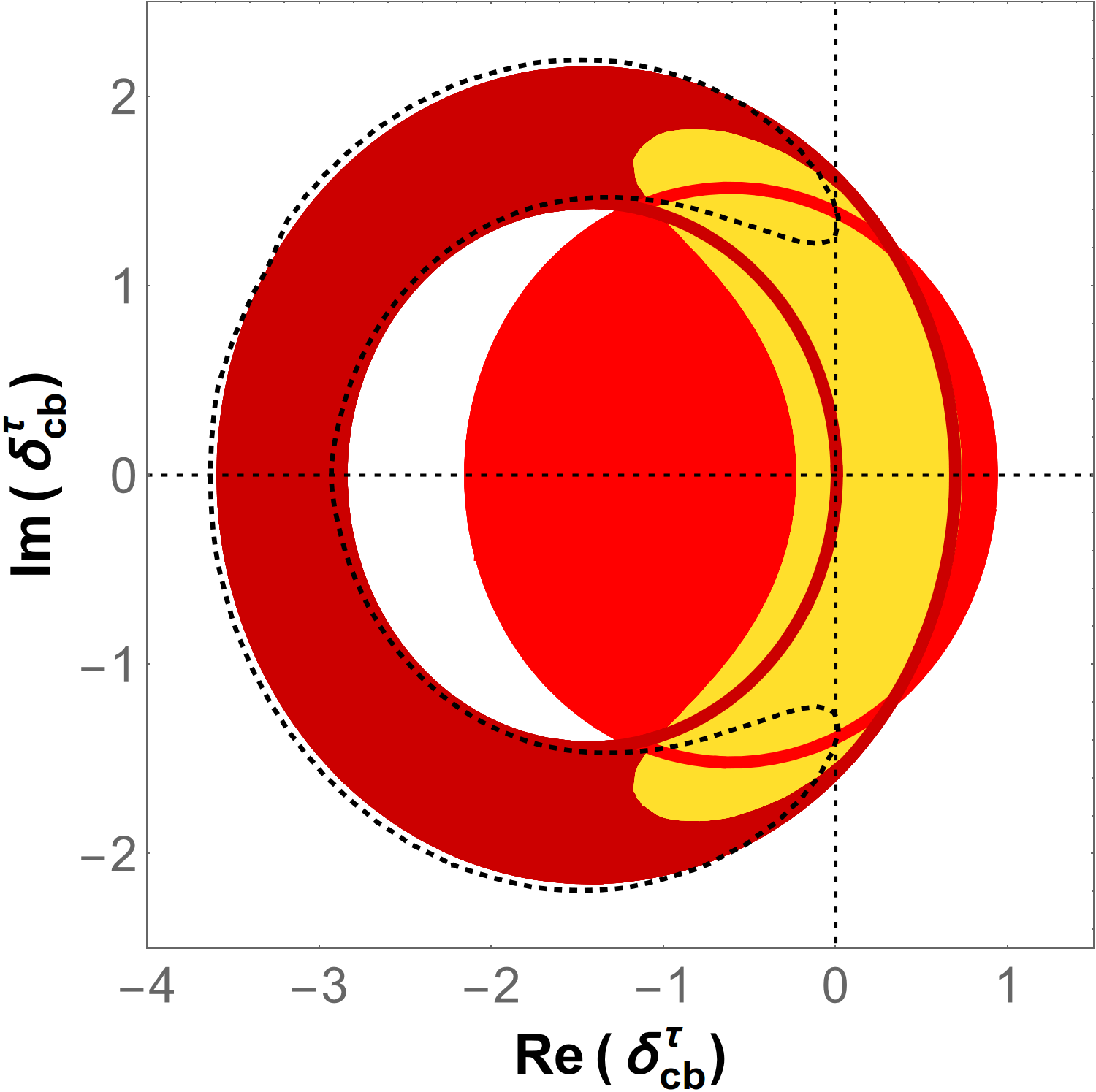}\qquad\qquad
\includegraphics[width=7.7cm,height=7.7cm]{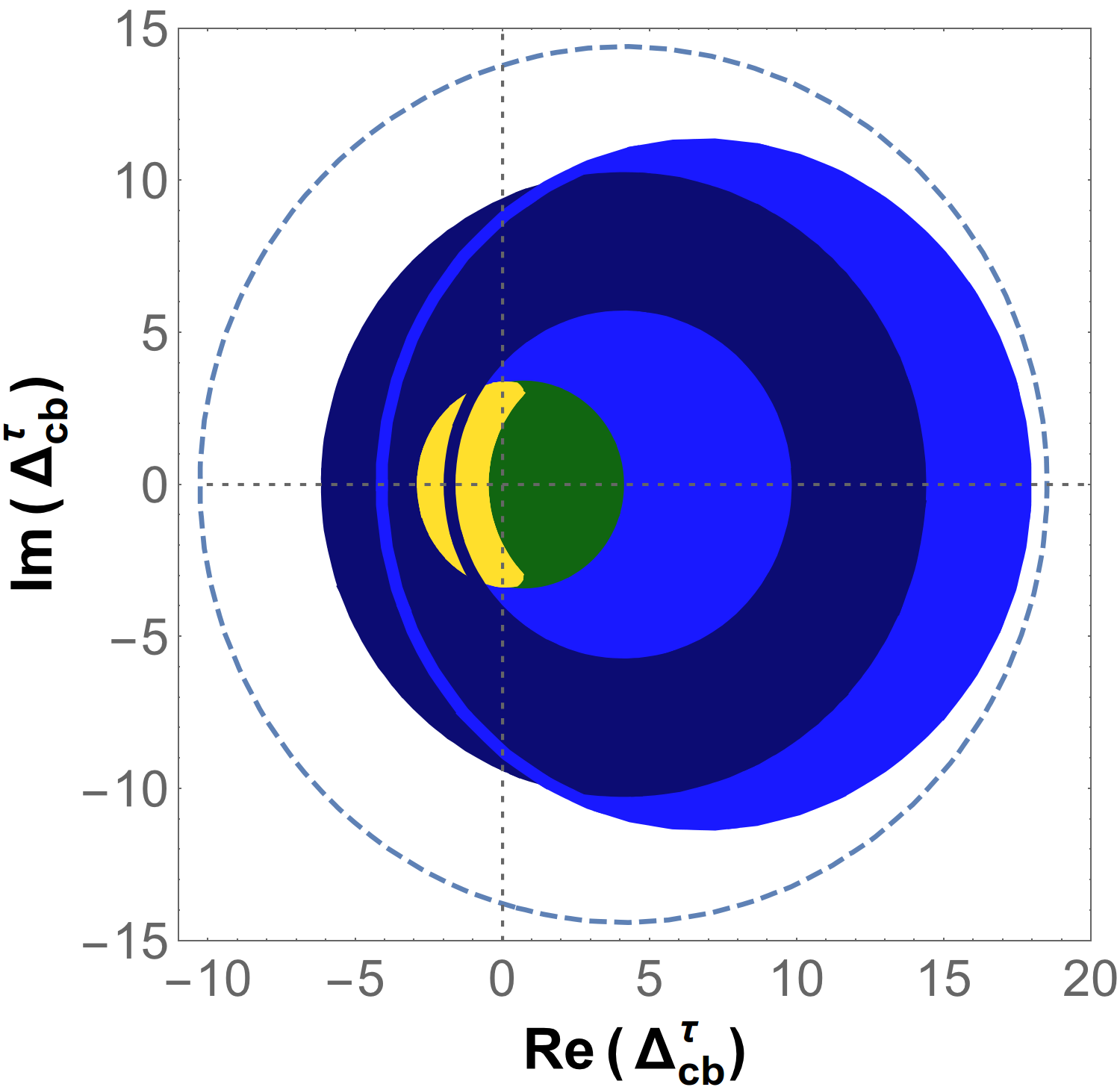}
\caption{\it \small Model-independent fits in the complex $\delta_{cb}^\tau$- (left) and $\Delta_{cb}^\tau$-planes (right). The dark
rings stem from \RDDs, the lighter discs from the shape information of the $q^2$-distributions of $\BtoDDs$, the dark green disc from
the indirect bound on $\mathrm{Br}(B_c\to\tau\nu)$ (see text), and the dashed contour in the right plot encloses the allowed region
from $A_\lambda(D^*)$. The yellow areas represent the global fit in each sector, while the dotted contour in the left plot encloses the
allowed region from a fit to \RDDs{} together with $R(\tau)$ in scenario S2, see text. All coloured areas correspond to $95\%$~CL regions, only the dashed contour to $68\%$~CL.
\label{fig:const}}
\end{figure*}

In Fig.~\ref{fig:const} we show the fit results for $B\to D\tau\nu$ data in the complex $\delta_{cb}^{\tau}$ plane (left), and the
$B\to D^*\tau\nu$ data together with the constraint from the total $B_c$ width $\Gamma_{B_c}$ in the complex $\Delta_{cb}^{\tau}$ plane
(right). For the $B\to D\tau\nu$ data we find that the $q^2$-distribution selects a part of the $R(D)$ ring that is closer to zero; its
preferred central value has a negative real part, opposite to the one from $R(D)$, rendering the combination well consistent with
the SM at $95\%$~CL.

For the $B\to D^*\tau\nu$ data, the differential distribution tends to exclude a part of the side of the $R(D^*)$ ring that is closer to zero, while extending
over the full ring on the other side. An important role is played by the constraint from the total $B_c$ width: it excludes a large
part of the parameter space preferred by the $R(D^*)$ measurement, including the second real solution in the complex $\Delta_{cb}^\tau$
plane, which was already discussed previously to be a highly fine-tuned solution~\cite{Celis:2012dk}.\footnote{Note that even
allowing for this mode to saturate the total rate, which is already contradicted by experiment, would still exclude the second real solution.} Specifically, it restricts the maximally allowed $R(D^*)$ values quite strongly. Nevertheless a consistent solution can be found for the available data, with the combined fit closer to the SM, and preferring large values for the branching ratio of $B_c\to\tau\nu$, see the discussion below.
The $\tau$ polarization asymmetry does presently not impose a further constraint on these couplings; its $68\%$~CL contour is shown for
completeness.

\begin{figure}
\includegraphics[width=7.2cm,height=7.2cm]{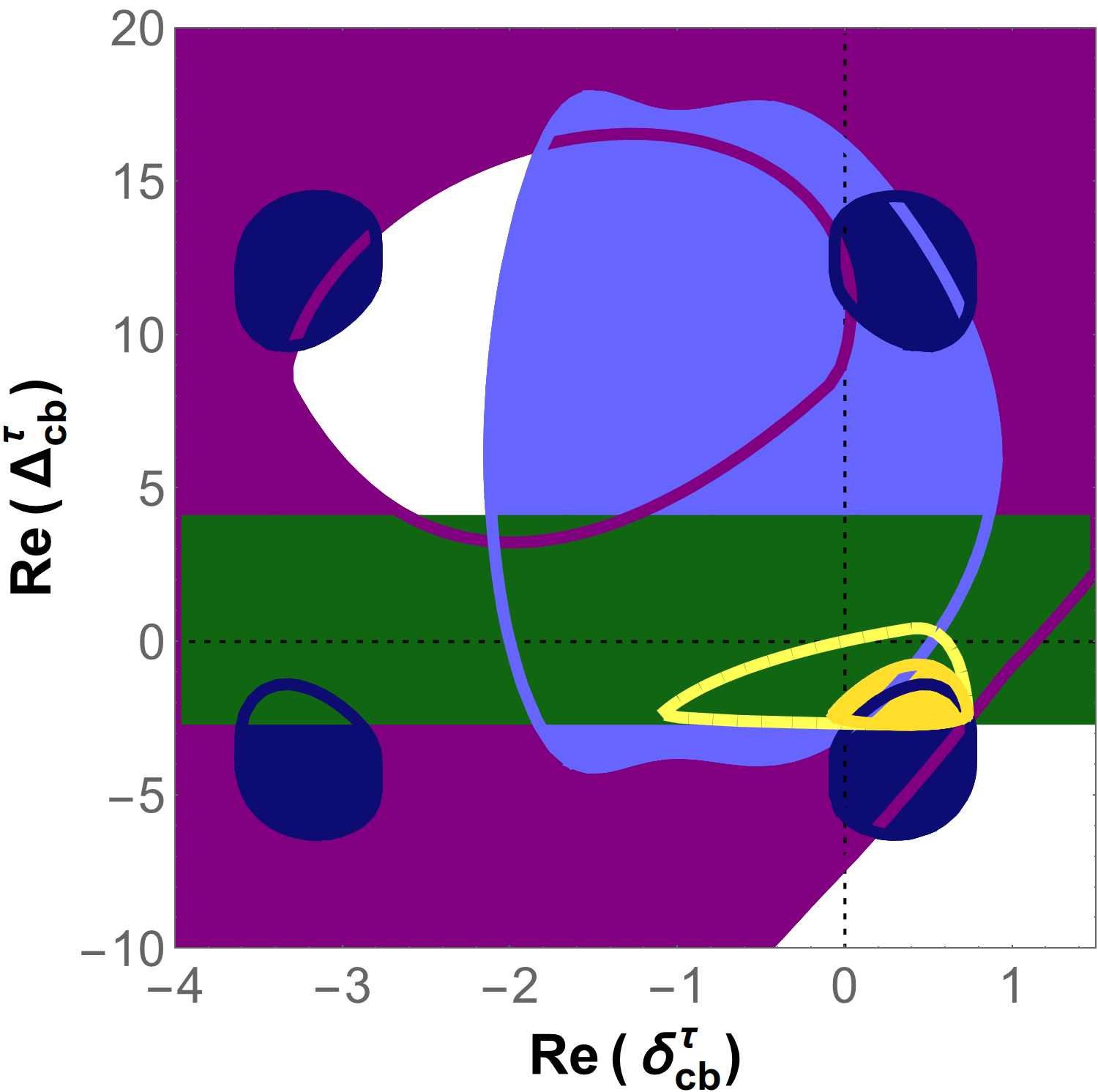}
\caption{\label{fig::deltareal} \it \small Constraints from \RDDs{} (dark blue), $R(X_c)$ (purple), the differential distributions in $\BtoDDs$ (light blue), and $\Gamma_{B_c}$ (dark green) in the $\delta_{cb}^\tau-\Delta_{cb}^\tau$-plane, assuming real couplings. The global fit is shown in dark yellow, while the light yellow contour shows how the global fit area extends when complex couplings are
allowed. All constraints are shown at $95\%$~CL.
}
\end{figure}

The \RDDs-rings in the complex $\delta_{cb}^\tau$- and $\Delta_{cb}^\tau$-planes yield four solutions when these parameters are chosen
to be real, as shown in Fig.~\ref{fig::deltareal}. The differential distributions exclude two of these solutions very clearly. A third
solution is excluded by the total $B_c$ width $\Gamma_{B_c}$, leaving an unambiguous solution, which shows however some tension with
the differential distributions and $\Gamma_{B_c}$, thereby shifting the global fit to lower values of $|\Delta_{cb}^{\tau}|$. $R(X_c)$
is seen to prefer smaller values for $|\Delta_{cb}^\tau+\delta_{cb}^\tau|$, but this constraint is shown here only for comparison and
is not included in the global fit.

The overall $\chi^2$ assuming real couplings does not increase compared to the general complex case, see Table~\ref{tab::chi2}, in
agreement with Fig.~\ref{fig:const}, where these imaginary parts are seen to be well compatible with zero. This is largely due to the
fact that the imaginary part enters the considered observables only quadratically, while the real part enters linearly. Improved
measurements of the included observables could nevertheless provide sensitivity on the imaginary part, since for instance the
constraints in form of disks from the distributions will turn into rings, yielding a potential non-trivial overlap with the ones from
\RDDs.

\begin{figure*}[htb]
\centering
\includegraphics[width=8cm,height=6cm]{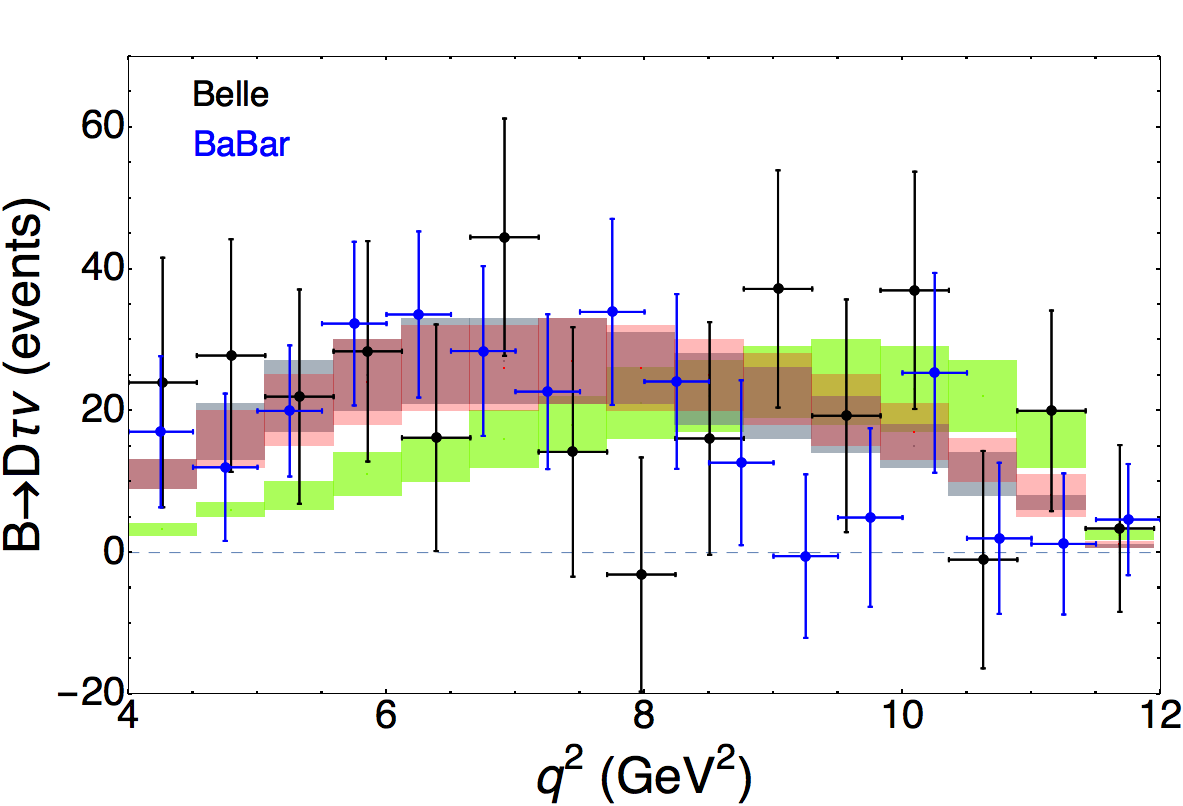}\qquad\qquad
\includegraphics[width=8cm,height=6cm]{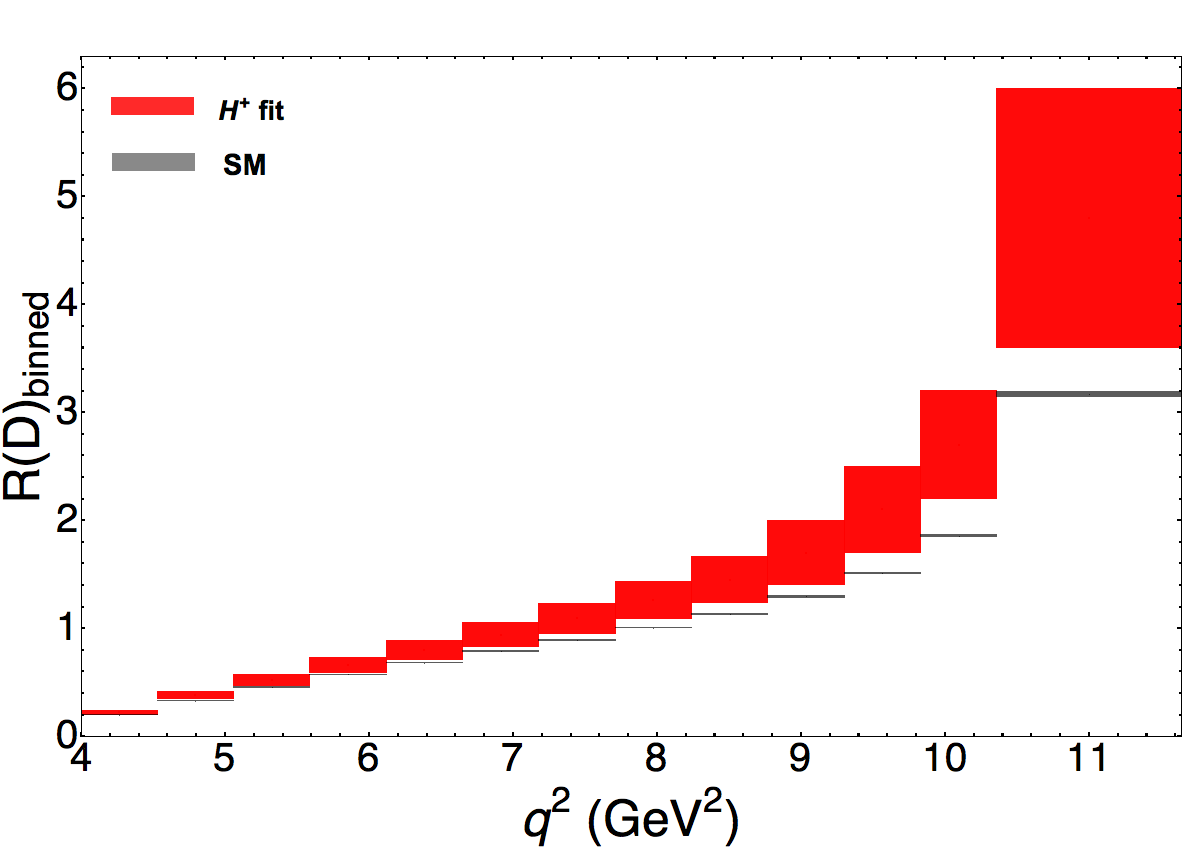}
\caption{ \it \small Left-panel: Measured differential distributions in $B \to D \tau \nu$ by BaBar and Belle, given as efficiency-corrected number of events as a function of the squared lepton invariant mass $q^2$. The $1\sigma$ ranges obtained from
the model-independent fit of $R(D)$ and the $q^2$ distribution are shown as solid-red bands. The result of a SM fit (excluding \RDDs)
is shown as solid-grey bands. The prediction for regions of the NP parameter space allowed by \RDDs, but excluded by the shape
information are shown as solid-green bands. Note that the BaBar data-points have been re-scaled by the relative normalization factor
obtained in the fit to have the same scale as the one from Belle. Right-panel: The $q^2$-binned SM prediction for $R(D)$, see Eq.~\eqref{eq::RDDsbinned}, and result from the fit including the scalar contribution.
\label{fig:diffdist}
}
\end{figure*}

\begin{figure*}[htb]
\centering
\includegraphics[width=8cm,height=6cm]{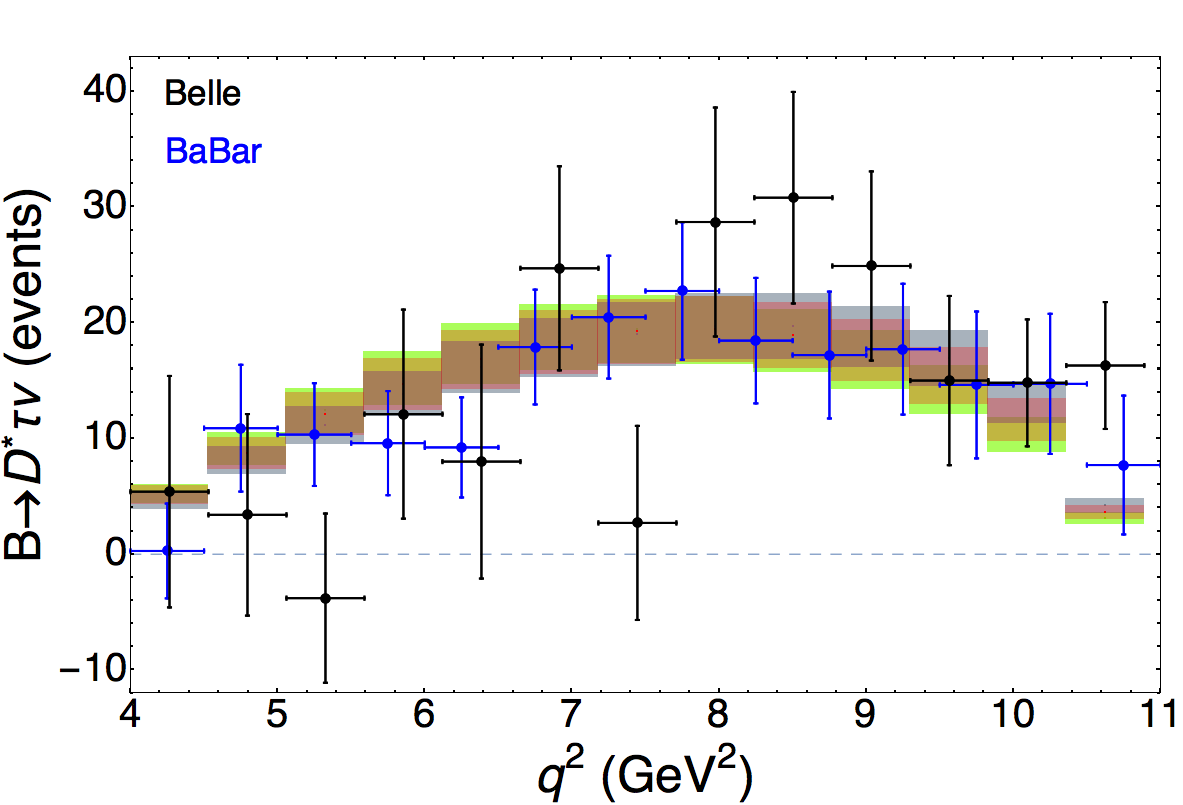}\qquad\qquad
\includegraphics[width=8cm,height=6cm]{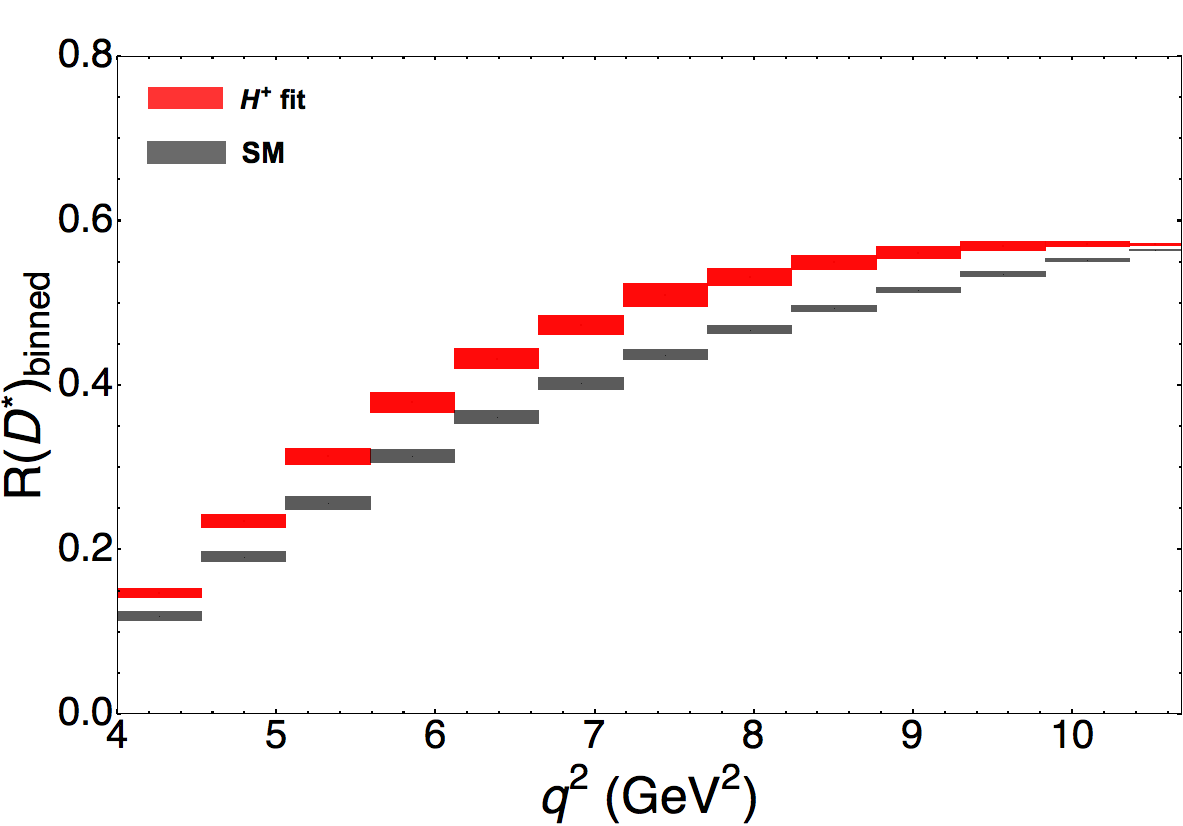}
\caption{ \it \small The caption is the same as in Fig.~\ref{fig:diffdist} but for $B \to D^* \tau \nu$.
}\label{fig:diffdistII}
\end{figure*}

To analyze the differential distributions in more detail, we show them in Figs.~\ref{fig:diffdist} and \ref{fig:diffdistII} on the left together with a model-independent NP fit (including \RDDs, red), a SM fit (excluding \RDDs, grey) and, for illustration, the NP prediction for values of $\delta_{cb}^\tau$ ($\Delta_{cb}^\tau$) that are allowed by \RDDs, but excluded by the shape information (green). Note again that the overall normalization for all four distributions, \emph{i.e.} the relations between yields and branching ratios,
are left free in the fits in order to decouple the information from the $q^2$ shapes from that of \RDDs, making a fit necessary also for the SM to fix them. The different normalization is also why the SM and NP distributions seem rather similar, although they correspond to very different physical pictures. The predicted $q^2$ distributions for $B\to D^{(*)}\tau\nu$ from the fit are given in Table~\ref{tab:distBelle}, with the normalization corresponding to the Belle data. In Figs.~\ref{fig:diffdist} and \ref{fig:diffdistII} on the right we show predictions for the $q^2$ spectrum of \RDDs{} from the model-independent NP fit and within the SM (without fitting). The numerical values for the $q^2$ spectrum of \RDDs{} shown in these figures are given in Table~\ref{tab:distRDRDstar}.

\begin{table}\begin{center}
\caption{\it \small Predicted $q^2$ distributions for \RDDs{} from the model-independent fit to $b\to c\tau\nu$ data and within the SM. }
\vspace{0.2cm}
\doublerulesep 0.8pt \tabcolsep 0.05in
\begin{tabular}{|c|c|c|c|c|}
\hline\rowcolor{RGray}
$q^2$ (GeV$^2$) & $R(D)|_{\rm fit}$ & $R(D)|_{\rm SM}$ & $R(D^*)|_{\rm fit}$  & $R(D^*)|_{\rm SM}$ \\
\hline 
$4.0-4.53$    & $0.22(2)$  & $0.199(1)$  & $[0.141,0.153]$ & $0.119(5)$\\\rowcolor{Gray}
$4.53-5.07$   & $0.38(3)$  & $0.330(1)$  & $[0.227,0.243]$ & $0.191(6)$\\
$5.07-5.6$    & $0.52(5)$  & $0.455(1)$  & $[0.303,0.323]$ & $0.256(8)$\\ \rowcolor{Gray}
$5.6-6.13$    & $0.66(7)$  & $0.571(2)$  & $[0.367,0.391]$ & $0.314(8)$\\
$6.13-6.67$   & $0.80(9)$  & $0.680(2)$  & $[0.420,0.444]$ & $0.361(8)$\\ \rowcolor{Gray}
$6.67-7.2$    & $0.94(11)$  & $0.786(3)$  & $[0.461,0.485]$ & $0.402(7)$ \\
$7.2-7.73$    & $1.09(14)$  & $0.892(3)$  & $[0.495,0.523]$ & $0.437(6)$\\ \rowcolor{Gray}
$7.73-8.27$   & $1.26(17)$    & $1.006(4)$  & $[0.521,0.541]$ & $0.467(5)$\\
$8.27-8.8$    & $1.45(21)$    & $1.135(5)$  & $[0.540,0.558]$ & $0.493(4)$\\ \rowcolor{Gray}
$8.8-9.33$    & $1.7(3)$    & $1.294(6)$  & $[0.554,0.568]$ & $0.516(3)$\\
$9.33-9.86$   & $2.1(4)$    & $1.513(7)$  & $[0.563,0.575]$ & $0.535(3)$\\ \rowcolor{Gray}
$9.86-10.4$   & $2.7(5)$    & $1.86(1)$  & $[0.568,0.574]$ & $0.552(2)$\\
$10.4-12.0$   & $4.8(1.2)$    & $3.17(2)$  & $[0.570,0.572]$ & $0.564(1)$\\ \rowcolor{Gray}
\hline
\end{tabular}
\label{tab:distRDRDstar}
\end{center}
\end{table}

As can be observed from these fits, the distributions available so far allow for sizeable NP contributions, while at the same time being compatible with the SM predictions, in accordance with the fits shown in Fig.~\ref{fig:const} and the analyses in Refs.~\cite{Lees:2013uzd,Huschle:2015rga}. On the other hand, the second NP distribution (green) is visibly different from the other two and in
clear tension with the data (especially for $B \to D \tau \nu$). For large NP contributions relative to the SM ones, the distributions
change due to kinematic effects. However, the region selected in $B\to D\tau\nu$ is safe from such large effects: the sharp drop
observed in Ref.~\cite{Lees:2013uzd} occurs for $\tan\beta/M_{H^\pm}\sim0.4~{\rm GeV}^{-1}$, which corresponds to
$\delta_{cb}^\tau\sim-2.4$, being far away from the global fit region. Therefore the global fit should be unaffected by this. For $B\to
D^*\tau\nu$, this effect is not very significant anywhere, so that also in this case our fit seems to be reliable. Regarding the
predictions for the differential distributions of \RDDs, clearly SM and NP are much easier to be disentangled from each other, since
the normalization factors cancel. Upcoming improved measurements of these distributions with more events will be particularly helpful to check
if the observed excess is due to a scalar contribution.

\begin{figure}[ht]
\centering{
\includegraphics[width=7.5cm,height=7.5cm]{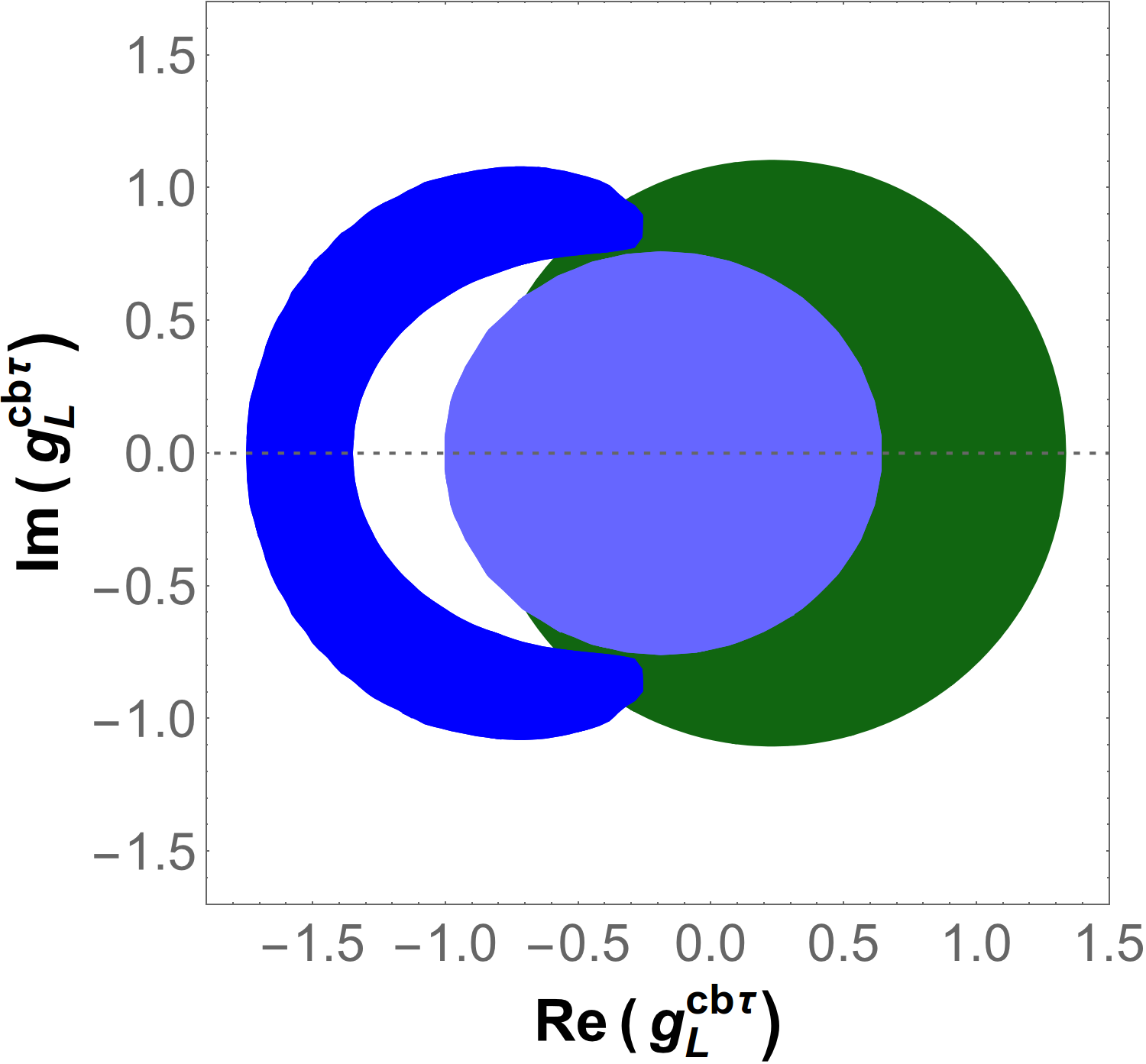}
\newline\vspace{-3ex}
\caption{ \it \small Constraints from \RDDs{} (blue), the differential distributions in $\BtoDDs$ (light blue) and $\Gamma_{B_c}$ (dark green) on the coefficient $g_{L}^{cb\tau}$ at $95\%$~CL, assuming $g_R^{cb\tau}$ to be zero.}
\label{fig:gLcons}
}
\end{figure}

We finish this model-independent analysis by discussing two sub-scenarios in which only one of the two couplings $g_{L,R}^{cb\tau}$ is present. For $g_L^{cb\tau}$ this has been observed as a possible solution to explain $R(D)$ and $R(D^*)$ in Ref.~\cite{Crivellin:2012ye}, and we confirm this including the new data. However, as illustrated in Fig.~\ref{fig:gLcons}, this scenario is in tension with the measured differential distributions as well as the total $B_c$ width. The resulting global fit remains better than the SM one, but worse than in scenario S1, both with real and complex couplings, see Table~\ref{tab::chi2} in Appendix~\ref{fitdetl}.  Especially the option of having a real $g_L^{cb\tau}$ as the common explanation for $R(D)$ and $R(D^*)$, which has been reiterated recently in Ref.~\cite{Crivellin:2015hha}, is highly disfavoured; the examples for excluded distributions shown in Figs.~\ref{fig:diffdist} and \ref{fig:diffdistII} belong exactly to this class of solutions.

The presence of $g_R^{cb\tau}$ alone does improve the fit to \RDDs{} compared to the SM one, but does not yield a good fit. Pursuing this option a bit further anyway, also in this case the situation is worsened by the differential distributions, although the minimal $\chi^2$ of the combination is similar to the one with $g_L^{cb\tau}$, only indicating less tension between differential distributions and \RDDs.

Adding both contributions simultaneously, as we did above, yields a better result than in both of these two sub-scenarios. Note that
this option has been ignored in Ref.~\cite{Sakaki:2014sea}, leading to the incorrect statement that scalar contributions alone could not explain \RDDs{} together with the measured differential distributions.

Finally, it is worth mentioning that none of the scenarios with NFC improves the description of \RDDs{} over the SM case: the only scenario that could affect these observables sizably is the Type-II 2HDM, but the constraints from $R(D)$ and $R(D^*)$ contradict each other in this case.\footnote{For this statement to hold strictly the effect on the differential distributions has to be taken into account; however, the BaBar analysis~\cite{Lees:2012xj,Lees:2013uzd} indicates that it holds even then.}

\subsubsection{$\mathbf{\boldsymbol{b} \to \boldsymbol{u \tau \nu}}$}

The semitauonic $b\to u$ transitions are less explored experimentally, given their additional suppression by $|V_{ub}/V_{cb}|^2\sim
1\%$. We find a mild tension for the experimental value of $R(\tau)$ with respect to the SM prediction, of about $1.8\sigma$, see
Table~\ref{TAB:EXP}. The measurement of $B\to\pi\tau\nu$ is not significant yet, and well compatible with the SM prediction within the
large uncertainties. Clearly both quantities are compatible with the SM as well as sizable scalar NP contributions, and cannot lead by
themselves to tensions within the model-independent scenario S1. However, the measured $R(\tau)$ already imposes a model-independent
correlation between $R(p) = \mathrm{Br}(\Lambda_b \to p \tau \nu)/\mathrm{Br}(\Lambda_b \to p \ell \nu)$ and $R(\pi)$, as discussed
below. Additionally, we observe that imposing a more specific flavour structure as in scenario S2 yields more stringent constraints,
discussed in the following.

\subsection{Universality of $\mathbf{\boldsymbol{b} \to \boldsymbol{c}}$ and $\mathbf{\boldsymbol{b} \to \boldsymbol{u}}$ -- S2\label{ssc}}

Assuming the flavour structure described in Eq.~\eqref{eq:rel} (S2), we obtain a more predictive scenario. Specifically, we can analyze the compatibility of $b\to c \tau \nu$ and $b \to u  \tau \nu$ data with a concrete assumption about the flavour structure of the underlying theory; this scenario remained viable after the BaBar result~\cite{Lees:2012xj}, see Ref.~\cite{Celis:2012dk}. However, taking into account all present data, the inclusion of $R(\tau)$ worsens the minimal $\chi^2$ significantly, $\Delta \chi^2_{\rm min}\approx 5$. The reason is that, while $R(D)$, $R(D^*)$ and $R(\tau)$ can be fitted simultaneously, $R(\tau)$ selects a region in the parameter space that is in
tension with the differential distribution of $B\to D\tau\nu$, as displayed in Fig.~\ref{fig:const} on the left as the dotted contour. Stated differently, the prediction for $R(D^*)$ excluding its experimental value, but including $R(\tau)$ is even smaller than the fitted value in the global $b\to c\tau\nu$ fit, preferring values below $0.28$.

\subsection{Differentiation between models}

In this subsection we investigate how additional measurements of $b\to (u,c)\tau \nu$ transitions can help to distinguish not only
between the SM and NP, but also between different NP scenarios. On the one hand, this is possible by fitting different models to the
available data, which yields different ranges and correlations between observables. On the other hand, in a given NP model, one can
construct combinations of observables in which the NP contributions cancel, such that the corresponding quantities can be predicted
independently of the NP considered. The operators in Eq.~\eqref{eq:Lag}, for instance, affect the polarization of the final-state
particles in a particular way, making it possible to distinguish the scalar effects from other dynamical scenarios; while the SM $W^-$
boson couples only to left-handed $\tau^-$ leptons, a charged-scalar would couple to $\tau^-$ leptons of the opposite chirality, and
would not enter in helicity amplitudes with a transversely polarized $D^*$ meson. Specifically, the following quantities remain
SM-like~\cite{Celis:2012dk}:
\begin{equation}\label{eq:relation2}
X_1(D^{*})=R(D^{*})-R_L(D^*)\,,
\end{equation}
where $R_L(D^{*})$ represents the decay rate for $B\to D^* \tau \nu$ normalized by the light lepton modes for longitudinally polarized $D^*$ mesons, see Ref.~\cite{Celis:2012dk} for the explicit expression,
and
\begin{equation}\label{eq:relation1}
X_2(D^{(*)})=R(D^{(*)})\left[A_\lambda(D^{(*)})+1\right]\,,
\end{equation}
which is built with the $\tau$ polarization asymmetry $A_\lambda(D^{(*)})$~\cite{Celis:2012dk}. The latter relation can also be generalized to semitauonic $\Lambda_b$ decays.

On the other hand, the scenario where the dominant NP effects in $b \to c \tau \nu$ have the same Lorentz structure as that of the SM
operator~\cite{Bhattacharya:2014wla,Greljo:2015mma,Calibbi:2015kma,Boucenna:2016wpr,Boucenna:2016qad}, parametrized as
\be \label{eq:leffgl}
\mathcal{L}_{\rm{eff}} = -\frac{4 G_F  V_{cb}  }{\sqrt{2}} g_{V_L} ( \bar c \gamma_{\mu}   \mathcal{P}_L b)  (\bar \tau
\gamma^{\mu} \mathcal{P}_L   \nu )   + \mathrm{h.c.}\,,
\ee
affects universally all ratios
\begin{equation}\label{eq:relation2b}
\hat R(X)\equiv \left.R(X)/R(X)\right|_{\rm SM}\,,
\end{equation}
and leaves unaffected all branching fractions that are normalized to quantities with the same transition, like the $\tau$ polarization
asymmetry, or double ratios like
\begin{equation}\label{eq:relation2c}
X_1^{V_L}(D^{*})=R_L(D^{*})/R(D^*)\,.
\end{equation}

\begin{figure}[hbt]
\centering{
\includegraphics[width=7.5cm]{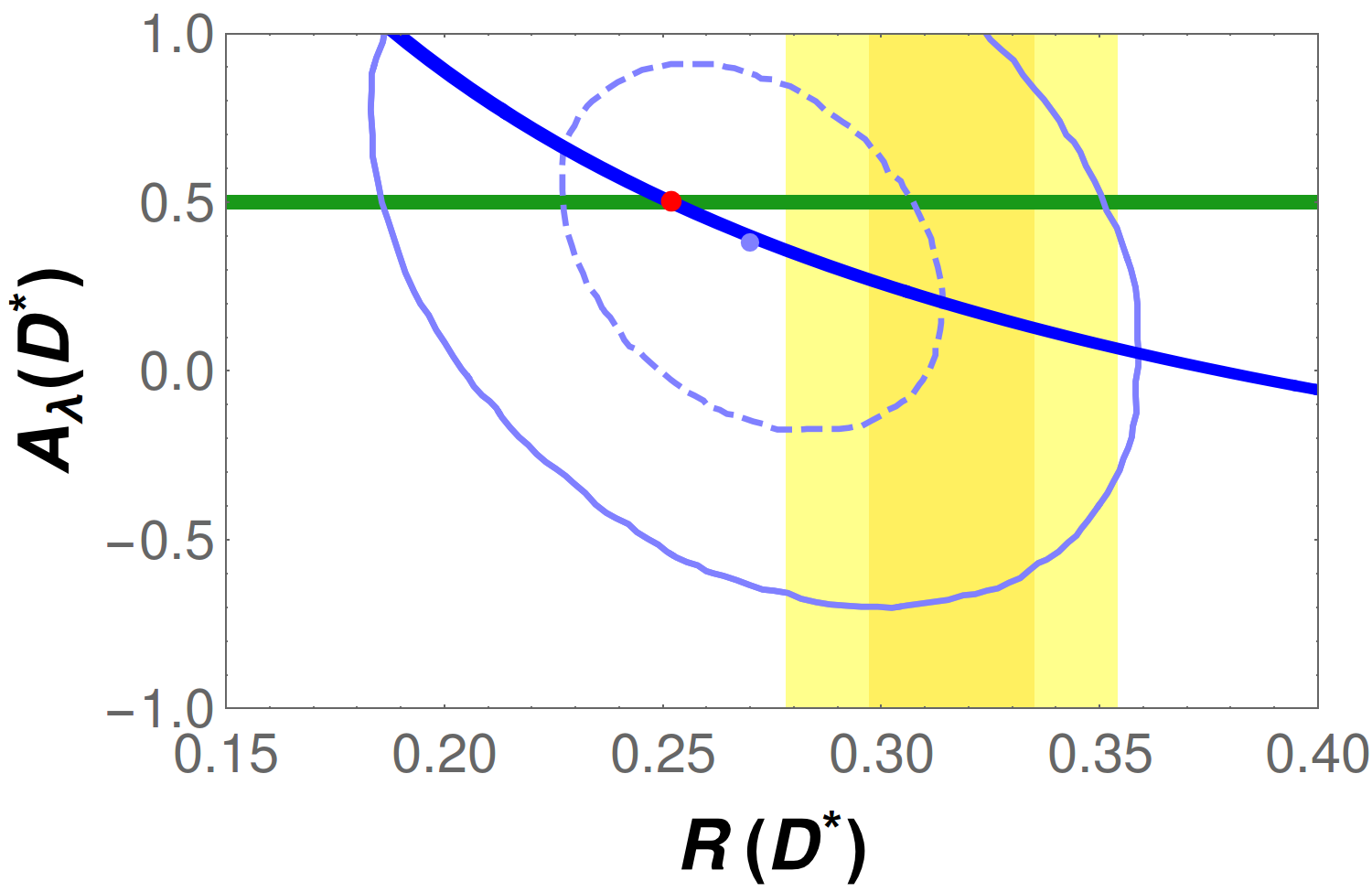}
\caption{\it \small Illustration of relation Eq.~\eqref{eq:relation1} (blue line) and experimental situation in the
$R(D^*)-A_\lambda(D^*)$ plane. The light blue dot and ellipses show the Belle measurement~\cite{Hirose:2016wfn} (central value and 1
and 2~$\sigma$ ellipses, respectively), the yellow band the average for $R(D^*)$ prior to this measurement, the green constant line
corresponds to the presence of only $g_{V_L}$, and the red dot to the SM prediction.
\label{fig:RDstarAlambda}
}
}
\end{figure}

\begin{figure}[hbt]
\centering{
\includegraphics[width=7.5cm]{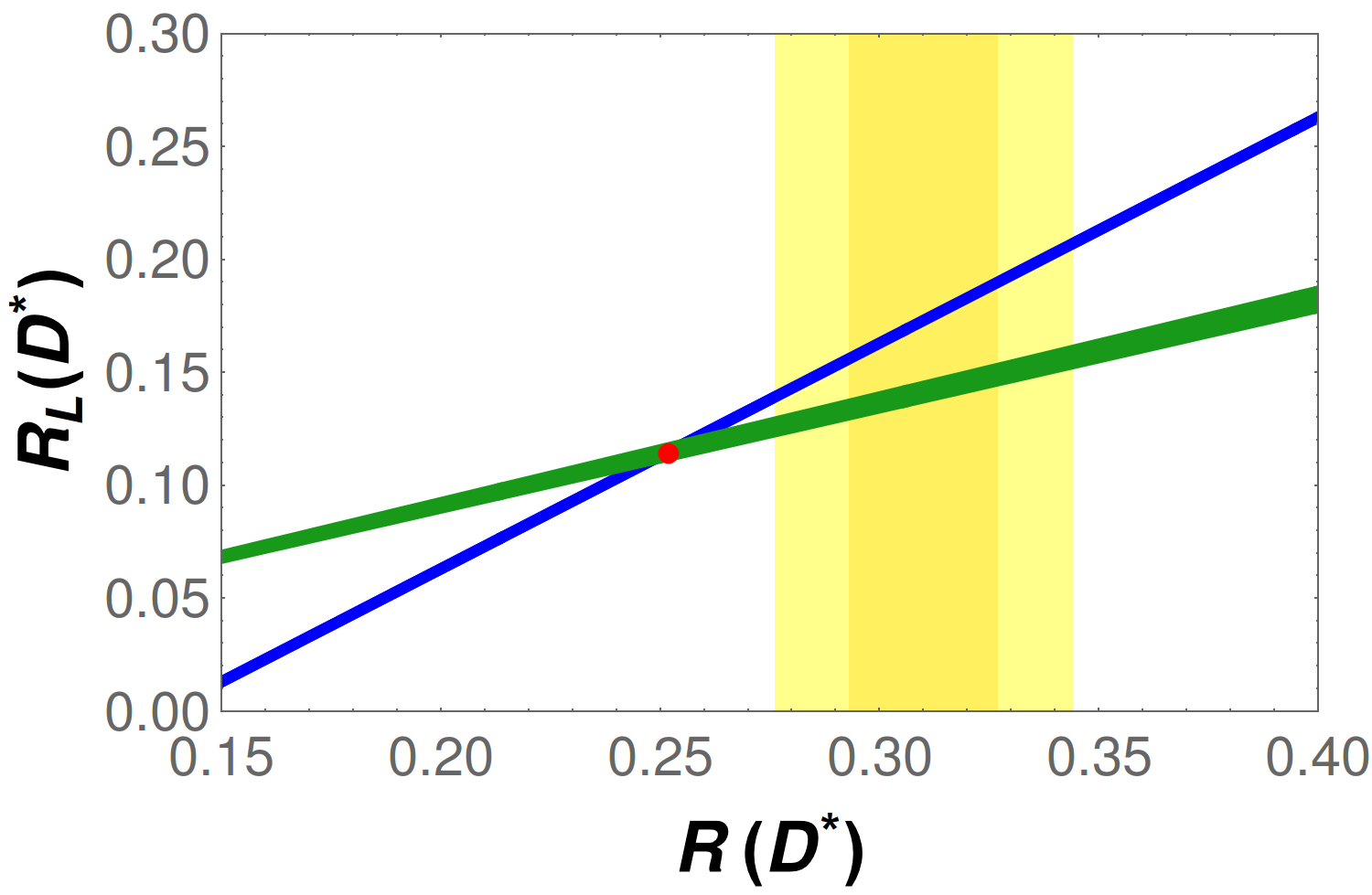}
\caption{\it \small Illustration of relations Eqs.~\eqref{eq:relation2} (blue line) and \eqref{eq:relation2b} (dark green line) in the $R(D^*)-R_L(D^*)$ plane. The yellow band shows the present average for $R(D^*)$, and the red dot corresponds to the SM prediction.
\label{fig:RDstarRLDstar}
}
}
\end{figure}

The observations regarding the polarization of the final-state particles are
illustrated in Figs.~\ref{fig:RDstarAlambda} and \ref{fig:RDstarRLDstar}. Although experimental uncertainties are still large for the $\tau$ polarization asymmetry and no measurement of the $D^*$ longitudinal polarization fraction has been performed yet, the potential of these observables to disentangle different dynamical scenarios is clear from these figures. Future measurements of $b \to c \tau \nu$ transitions performed at the LHCb and Belle II experiments can exploit these possibilities.

Another generic difference between the two NP scenarios is the relation between $R(D^*)$ and $\mathrm{Br}(B_c\to\tau\nu)$, already
discussed for the scalar case above: the $B_c$ branching ratio is very sensitive to charged-scalar effects, yielding large enhancements for the present central value of $R(D^*)$, while with SM-like couplings the enhancement is moderate. Since this mode is very difficult to
measure, the limit stems from the total width of the $B_c$ meson, see Appendix~\ref{app:diff}. As can be seen from
Fig.~\ref{fig:BRBchRDstar}, the present value for $R(D^*)$ shows some tension with the total $B_c$ width for scalar NP, while
there is no limit on the SM-like coupling. Because of this tension, the global fit for the SM-like coupling is slightly better
than the one with scalar NP, see Table~\ref{tab::chi2}. However, both scenarios still improve the fit significantly compared to the
SM.

\begin{figure}[hbt]
\centering{
\includegraphics[width=7.5cm]{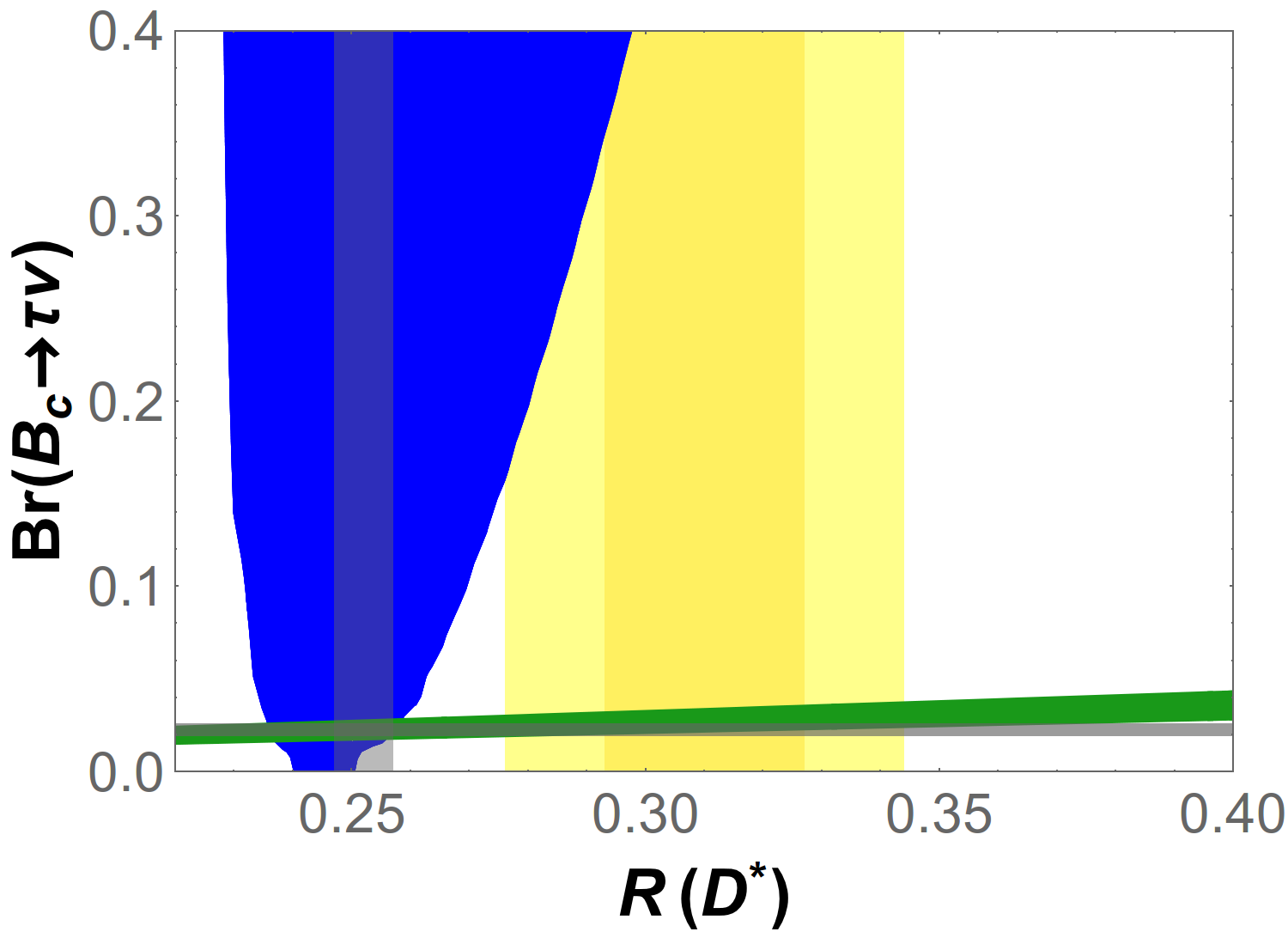}
\caption{\it \small $\mathrm{Br}(B_c\to \tau \nu)$ versus $R(D^*)$ in the SM (grey bands), scalar NP (blue area) and left-handed vector NP (dark green band). The yellow band shows the present average for $R(D^*)$.
\label{fig:BRBchRDstar}
}
}
\end{figure}

\begin{figure}[hbt]
\centering{
\includegraphics[width=7.5cm]{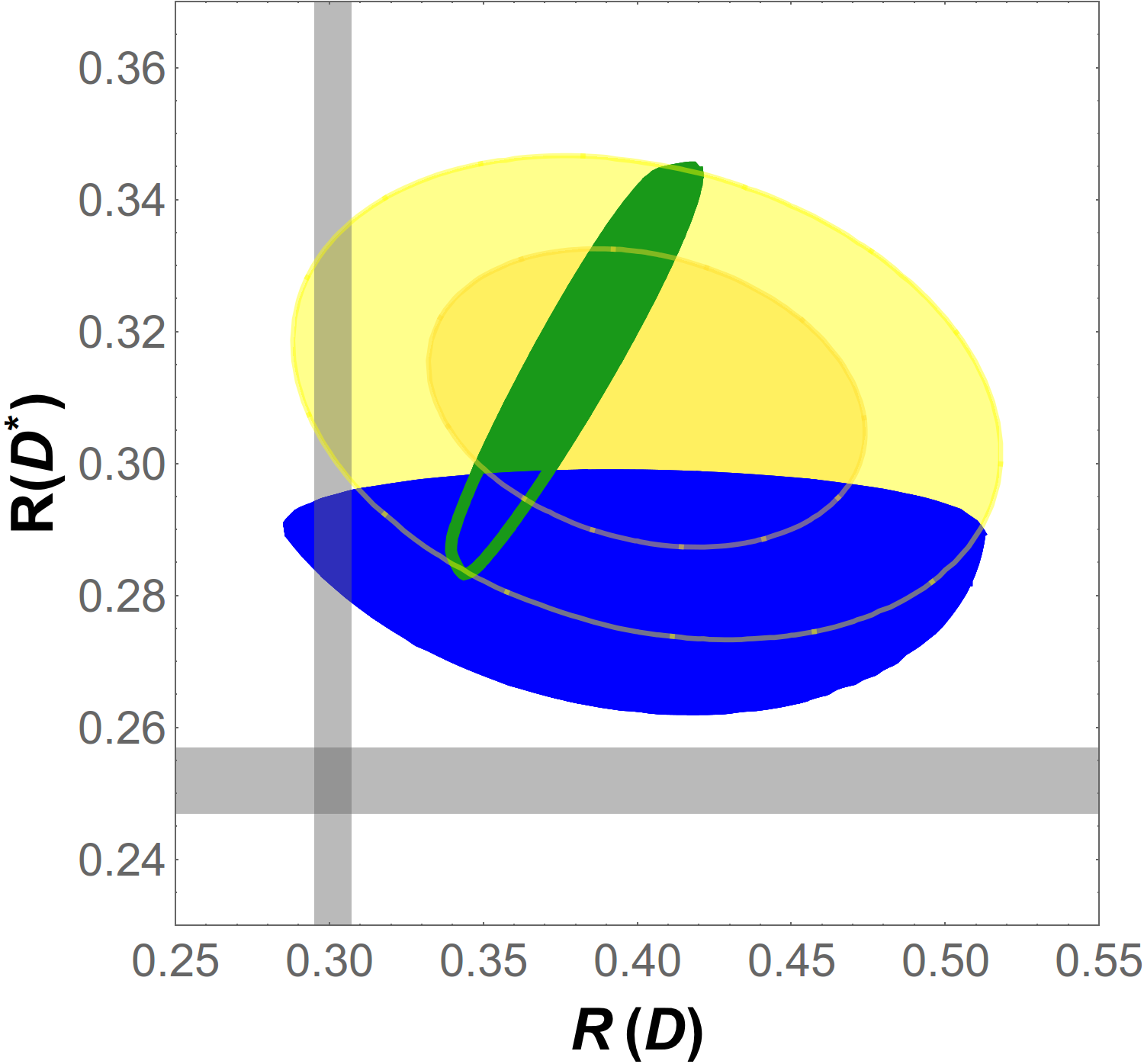}
\caption{\it \small Fit result for $R(D)$ versus $R(D^*)$ from a global fit with scalar operators (blue area) and with a left-handed
vector coupling (green area), together with the SM prediction (grey bands) and the experimental average (yellow ellipses). All areas
correspond to $95\%$~CL, only the dark yellow one to $68\%$~CL.
\label{fig::RDRDs}
}
}
\end{figure}

\begin{figure}[hbt]
\centering{
\includegraphics[width=7.5cm]{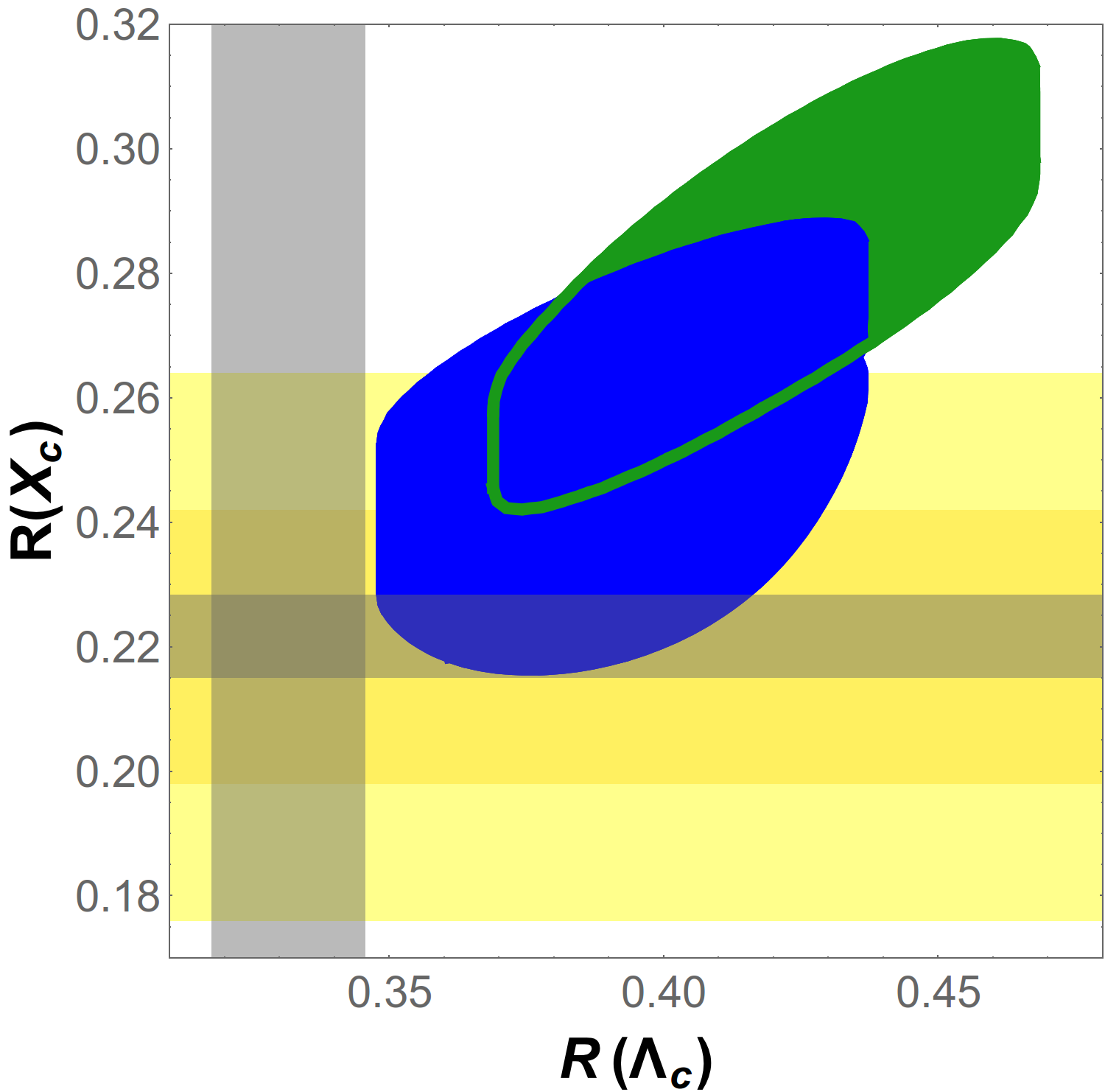}
\caption{\it \small Prediction for $R(X_c)$ versus $R(\Lambda_c)$ from a global fit with scalar operators (blue area), a global fit with a left-handed vector coupling (green area), together with the SM prediction (grey bands) and the $R(X_c)$ measurement by LEP (yellow bands). All bands correspond to $95\%$~CL, only the dark yellow one to $68\%$~CL.
\label{fig:RLambdacRXc}
}
}
\end{figure}

Figs.~\ref{fig::RDRDs} and \ref{fig:RLambdacRXc} show our fit results for some key observables with present data. The fit results for $R(D)$
and $R(D^*)$ in both NP scenarios are shown in Fig.~\ref{fig::RDRDs}; for scalar NP, this fit yields a range for $R(D^*)$ that is
larger than in the SM, but smaller than the present experimental central value, while with left-handed vector NP this value can be
reached for $R(D^*)$, but $R(D)$ is predicted to be smaller than the present experimental central value, due to the aforementioned
strong correlation $\hat R(D)=\hat R(D^*)$. Fig.~\ref{fig:RLambdacRXc} shows the predictions for $R(X_c)$ and $R(\Lambda_c) =
\mathrm{Br}(\Lambda_b \to \Lambda_c \tau \nu)/\mathrm{Br}(\Lambda_b \to \Lambda_c \ell \nu)$ from a global fit to the other $b\to
c\tau\nu$ observables in both NP scenarios; in both cases enhancements for these two observables are expected with respect to the SM
predictions. The predicted enhancements are larger in the case of a left-handed vector coupling, which is in slight tension with the
available measurement for $R(X_c)$. Again (more) precise measurements for these observables can help to distinguish the two NP
scenarios.

\begin{figure}[hbt]
\centering{
\includegraphics[width=7.5cm]{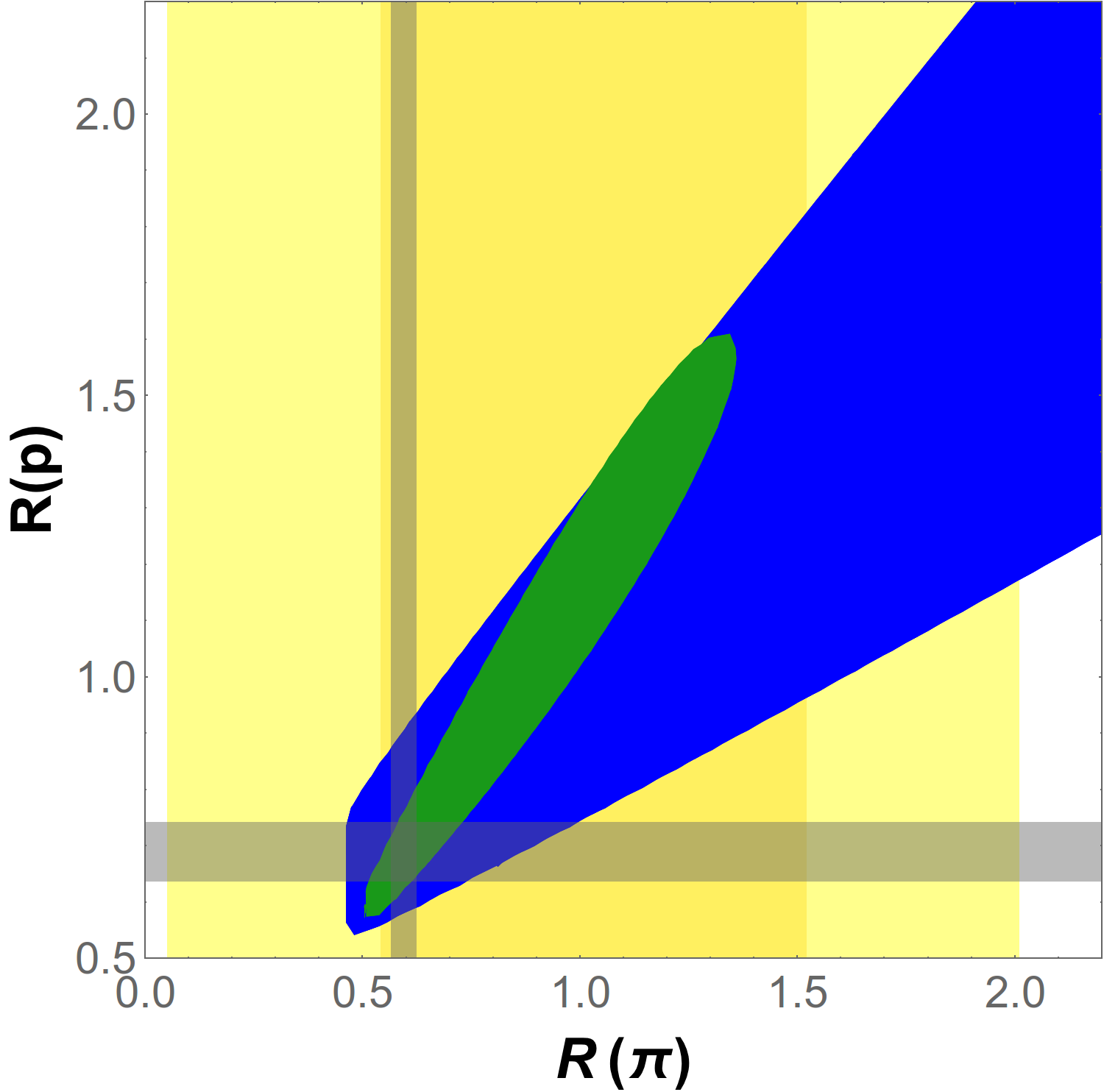}
\caption{\it \small Prediction for $R(p)$ versus $R(\pi)$ from a fit to $R(\tau)$ with scalar operators (blue area) and with a left-handed vector coupling (green area), together with the SM prediction (grey bands) and the $R(\pi)$ measurement by Belle (yellow bands). All bands correspond to $95\%$~CL, only the dark yellow one to $68\%$~CL.
\label{fig:RpiRp}
}
}
\end{figure}

Finally, considering the same NP structure as in Eq.~\eqref{eq:leffgl} for $b\to u \tau\nu$ transitions, we show in Fig.~\ref{fig:RpiRp} the correlation between the $b\to u\tau\nu$ observables $R(p)$ and $R(\pi)$ as predicted from the
available measurement of $R(\tau)$. Large enhancements as well as SM-like values are possible for both observables, given that
$R(\tau)$ is still compatible with the SM prediction. Furthermore, their correlation is again different in the two NP scenarios,
providing therefore a means to distinguish them in $b\to u\tau\nu$ transitions.

\section{Conclusions} \label{sec:conc}

We have performed a comprehensive analysis of current $b \to c (u) \tau \nu$ data in the presence of generic scalar
contributions, providing additionally predictions for $\Lambda_b \to \Lambda_c \tau \nu$, $B \to X_c \tau \nu$, $\Lambda_b \to p
\tau \nu$, and $B \to \pi \tau \nu$ from a global fit to the other $b\to c (u) \tau\nu$ observables.

We analyzed the possibility to alleviate the current tension between $b \to c \tau \nu$ data and the SM predictions, which is at the
level of about $4 \sigma$. Compared to the SM case, we find that scalar contributions with the simultaneous presence of both
left- and right-handed couplings to quarks can improve considerably the global fit of $R(D^{(*)})$ and the measured $q^2$ differential
distributions in $B \to D^{(*)} \tau \nu$. The indirect bound derived from the total $B_c$ width is also included in the analysis and
plays an important role by excluding a large part of the parameter space preferred by the $R(D^*)$ measurement, as shown in
Figs.~\ref{fig:const} and \ref{fig::deltareal}. As a result, an explanation of the tension with scalar contributions requires
values for $R(D^*)$ to be $1-2$ standard deviations smaller than the present experimental central value. Restricted scenarios with scalar couplings involving only left- or right-handed scalar couplings to quarks are found to be disfavoured by the $q^2$ differential distributions in $\BtoDDs$ and the total $B_c$ width.

Finally, we also discussed the possibility to disentangle scalar effects in $b \to c (u) \tau \nu$ transitions from other NP
scenarios, specifically the presence of only a left-handed vector current. Observables involving the polarization of the final $\tau$ lepton and the $D^*$ meson show strong correlations which can be predicted even in the presence of NP with high precision.
Furthermore, different patterns are predicted for decay modes like $\Lambda_b \to \Lambda_c \tau \nu$, $B \to X_c \tau \nu$, $\Lambda_b
\to p \tau \nu$, and $B \to \pi \tau \nu$. These findings can be further exploited by future measurements of $b \to c (u) \tau
\nu$ transitions at the LHCb and Belle II experiments.

\acknowledgments
We are grateful to Manuel F. Sevilla and Thomas Kuhr for providing details about the experimental measurements. The work of A.C. is supported by the Alexander von Humboldt Foundation.  A.C. is also grateful to the Mainz Institute for Theoretical Physics (MITP), the Universit\`a di Napoli Federico II and INFN for its hospitality and its partial support during the completion of this work.  The work of M.J. is financially supported by the ERC Advanced Grant project ``FLAVOUR'' (267104) and the DFG cluster of excellence ``Origin and Structure of the Universe''. The work of X.L. is supported by the
NNSFC~(Grant Nos. 11675061 and 11435003), by the SRF for ROCS, SEM, and by the self-determined research funds of CCNU from the colleges' basic research and operation of MOE~(CCNU15A02037).  The work of A.P. is supported by the Spanish Government and ERDF funds from the EU Commission [Grant FPA2014-53631-C2-1-P], by the Spanish Centro de Excelencia Severo Ochoa Programme [Grant SEV-2014-0398] and by the Generalitat Valenciana [Grant PrometeoII/2013/007]. The computations have partly been carried out on the computing facilities of the Computational Center for Particle and Astrophysics (C2PAP). This research was also supported by the Munich Institute for Astro- and Particle Physics (MIAPP)
of the DFG cluster of excellence ``Origin and Structure of the Universe''.

\appendix

\section{Hadronic input parameters and statistical treatment} \label{sec:inputs}

The hadronic input parameters used in our analysis are listed in Table~\ref{TAB:inputs}. For $R(D^*)$ we use the Caprini-Lellouch-Neubert (CLN) parametrization~\cite{Caprini:1997mu}. The corresponding form factor parameters are extracted from data~\cite{Amhis:2016xyh}, apart from the form factor ratio $R_3(1)$, which is obtained using a HQET relation to order $\alpha_s,1/m_{b,c}$~\cite{Falk:1992wt,Falk:1992ws,Neubert:1992tg} and enhancing the related uncertainty to account for higher-order effects~\cite{Fajfer:2012vx}. For $R(D)$, we use the recent determination of the $B\to D$ form factors in Ref.~\cite{Bigi:2016mdz},
employing the Boyd-Grinstein-Lebed parametrization~\cite{Boyd:1994tt,*Boyd:1997kz}. The relevant inputs for our $R(D)$ prediction are quoted in Tables 4 and 5 of Ref.~\cite{Bigi:2016mdz}; we use the results of the $N=2$ fit.

\begin{table}[ht]
\centering{
\caption{\label{TAB:inputs} \it \small  Input values for the relevant hadronic parameters, see text for details.  } } \vspace{0.2cm}
\doublerulesep 0.8pt \tabcolsep 0.02in
\small{   \begin{tabular}{ccc}\hline\hline   \rowcolor{RGray}
Parameter                                            & Value                                   &
Comment \\ \hline
\rowcolor{Gray}$f_{B_c}$                             & $(434\pm15)$~MeV                        &
\cite{Colquhoun:2015oha}\\
$f_{B_u}$                                            &
$(187.1\pm 4.2)~{\rm MeV}$          &
\cite{Rosner:2015wva}\\    \rowcolor{Gray}
$|V_{cb}|$                                           & $(40.5\pm1.5)\times 10^{-3}$ & \cite{Olive:2016xmw}  \\
$R_1(1)$                                             & $1.406 \pm 0.033 $                      &  \cite{Amhis:2016xyh}\\\rowcolor{Gray}
$R_2(1)$                                             & $0.853 \pm 0.020 $                      & \cite{Amhis:2016xyh}\\
$R_3(1)$                                             & $ 0.97 \pm 0.10 $                       &
\cite{Falk:1992wt,Falk:1992ws,Neubert:1992tg} \\\rowcolor{Gray}
$\rho^2$                                             & $1.207 \pm0.026 $                       &  \cite{Amhis:2016xyh}\\
\hline\hline
\end{tabular}}
\end{table}

We call the attention to recent works on the determination of the $B\to D^{(*)}$ form factors. In
Refs.~\cite{Lattice:2015rga,Na:2015kha} a model-independent parametrization of the form factors based on analyticity and
unitarity~\cite{Boyd:1994tt,*Boyd:1997kz} has been used, in this case avoiding the use of the CLN parametrization. The values obtained for $R(D)$ in these works are $R(D) = 0.299\; (11)$~\cite{Lattice:2015rga}
and $R(D) = 0.300\; (8)$~\cite{Na:2015kha}, using additionally experimental input from $B\to D\ell\nu$. Another recent work employs
perturbative QCD factorization and lattice QCD inputs to extract the relevant $B \to D^{(*)}$ form factors, finding $R(D) =
0.337^{+0.038}_{-0.037}$ and $R(D^*) = 0.269^{+0.021}_{-0.020}$~\cite{Fan:2015kna}. We note that these predictions are compatible with
ours at the $1\sigma$ level.  Finally, the possible pollution of $R(D^*)$ from $B^*$ pole contributions has been found to be
negligible~\cite{Kim:2016yth}.

\begin{figure}[t]
\centering
\includegraphics[width=8.cm,height=5.5cm]{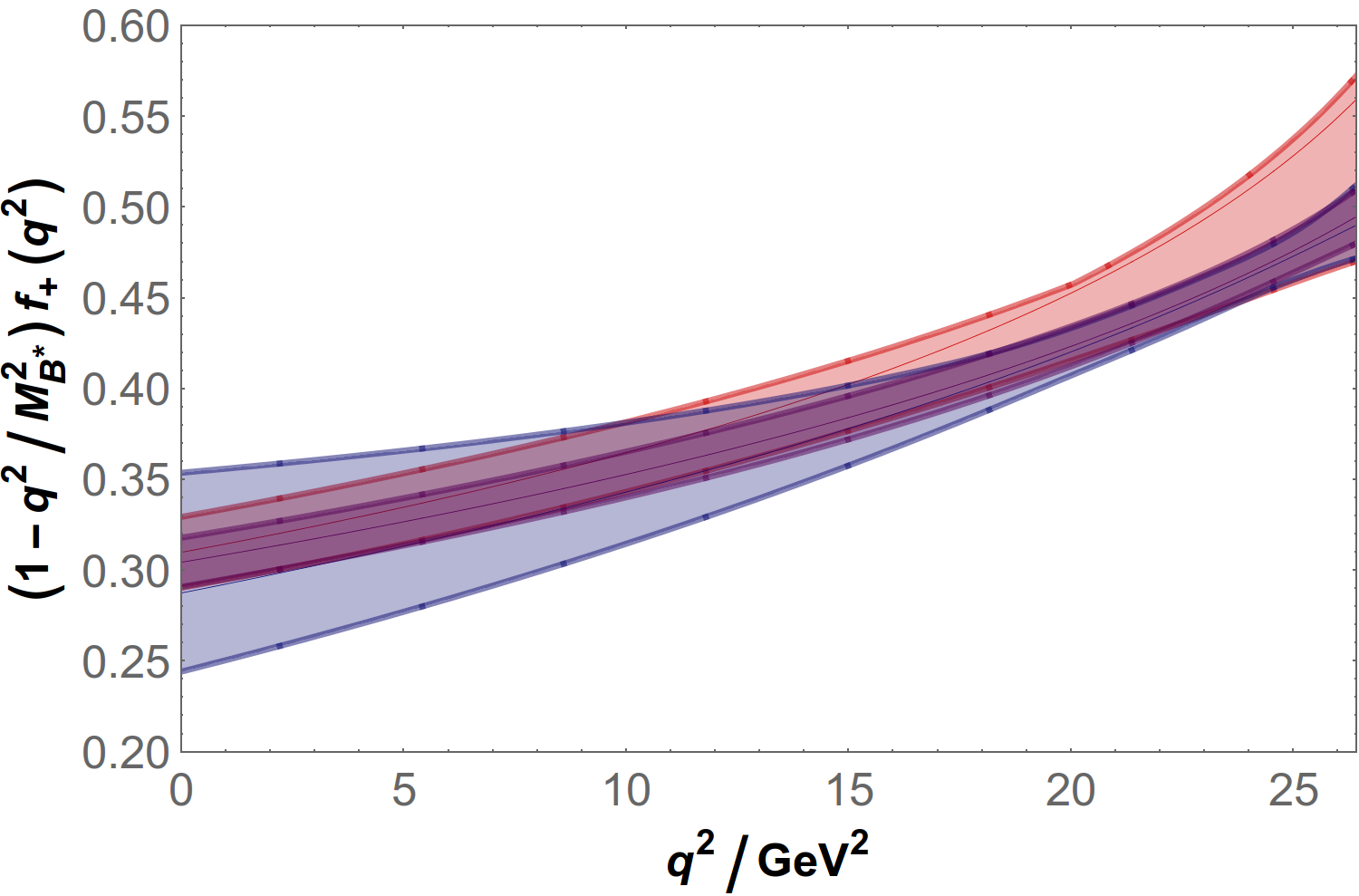}
\caption{ \it \small The form factor $f_+^{B\to\pi}(q^2)$ determined from lattice calculations at large values of $q^2$ (red),
LCSR calculations at small values of $q^2$ (blue), and their combination (purple), using the BCL parametrization over the
whole $q^2$ range.
\label{fig::BtopiFF}
}
\end{figure}

\begin{table*}[t]
\centering{
\caption{\label{tab::BtopiFF} \it \small  Form-factor parameters for $f_+^{B\to\pi}(q^2)$ and $f_0^{B\to\pi}(q^2)$ in the
Bourrely-Caprini-Lellouch (BCL) parametrization~\cite{Bourrely:2008za} with N=2 and N=3 fits, respectively. Note that the
coefficient $b_2$ has been eliminated by the constraint $f_0^{B\to\pi}(0)=f_+^{B\to\pi}(0)$.}  \vspace{0.2cm}\doublerulesep 0.8pt \tabcolsep 0.02in
\small{
\begin{tabular}{c c c c c c c}\hline\hline   \rowcolor{RGray}
Parameter  & $a_0$              & $a_1$             & $a_2$         & $b_0$             & $b_1$         & $b_3$\\\hline
Value      & $0.424\pm0.011$    & $-0.333\pm0.039$  & $-0.31\pm0.08$& $0.515\pm0.019$   & $-1.65\pm0.08$&
$5.0\pm0.9$\\\hline\rowcolor{RGray}
$a_0$      & 1                  & 0.19              & $-0.44$         & 0.04              & 0.13          & 0.09\\
$a_1$      & 0.19               & 1                 & $-0.50$         & 0.04              & 0.12          & 0.08\\\rowcolor{RGray}
$a_2$      & $-0.44$            & $-0.50$             & 1             & $-0.01$             & $-0.04$         & -0.02\\
$b_0$      & 0.04               & 0.04              & $-0.01$         & 1                 & $-0.03$         & -0.41\\\rowcolor{RGray}
$b_1$      & 0.13               & 0.12              & $-0.04$         & $-0.03$             & 1             & $-0.75$\\
$b_3$      & 0.09               & 0.08              & $-0.02$         & $-0.41$             & $-0.75$         & 1\\\rowcolor{RGray}
\hline\hline
\end{tabular}
}}
\end{table*}

For the $B\to\pi$ form factors necessary for the prediction of the $B \to \pi \tau \nu$ decay we proceed as follows: we use the information on the parameters of the vector form factor $f_+^{B\to\pi}(q^2)$ from two recent lattice calculations~\cite{Lattice:2015tia,Flynn:2015mha} at large values of $q^2$, as combined by FLAG~\cite{Aoki:2016frl}, together with the information from a recent light-cone sum rule (LCSR) calculation~\cite{Imsong:2014oqa} at small values of $q^2$, to obtain a reliable vector form factor over the whole $q^2$ range. In the same combination we use, for the lack of a combination by FLAG, the
results for the scalar form factor $f_0^{B\to\pi}(q^2)$ from Ref.~\cite{Lattice:2015tia},\footnote{These values are more precise
than the ones given in Ref.~\cite{Flynn:2015mha}, which are calculated from a subset of the same lattice ensembles, complicating a
simple combination. The recent calculation in Ref.~\cite{Colquhoun:2015mfa} only provides a value for $f_0^{B\to\pi}(q^2_{\rm max})$,
which is however about $2\sigma$ higher than the value implied by our form-factor fit.} imposing additionally the constraint
$f_0^{B\to\pi}(0)=f_+^{B\to\pi}(0)$ to eliminate the coefficient $b_2$, which introduces (small) correlations between the $a_i$ and
$b_i$ parameters. This combination works very well, see Fig.~\ref{fig::BtopiFF}. The resulting form factor parameters and their
correlations are given in Table~\ref{tab::BtopiFF}.

For the predictions of $\Lambda_b \to \Lambda_c \ell \nu$ and $\Lambda_b \to p \ell \nu$ decays we use the transition form factors
determined from lattice QCD~\cite{Detmold:2015aaa}.

Bounds on the parameter space are obtained using frequentist statistics and the ``Rfit'' treatment for theoretical uncertainties~\cite{Hocker:2001xe}. However, there is a very limited amount of quantities which receive large theory uncertainties that are difficult to quantify: these are mainly the form factor ratio $R_3(1)$ and the
coefficients $c_i$ in $B\to X_c\tau\nu$, to be discussed in the next subsection. These lead to ``flat'' uncertainties (in the sense of Rfit) which are quoted as a second uncertainty in Table~\ref{TAB:EXP} in the SM predictions. All nuisance parameters are kept
floating in the fits.

\section{Details on $\mathbf{\boldsymbol{b}\to \boldsymbol{c\tau\nu}}$ observables} \label{app:diff}

The $q^2$ distributions for $\BtoDDs$ decays measured by Belle~\cite{Huschle:2015rga} and BaBar~\cite{Lees:2013uzd} are given as
efficiency-corrected number of events and collected in Tables~\ref{tab:distBelle} and \ref{tab:distBaBar}. They are given for
$q^2\geq4~{\rm GeV}^2$ due to the experimental selection criteria~\cite{Lees:2013uzd,Huschle:2015rga}. The uncertainties given for the
individual bins only include the statistical ones. To account for the systematic uncertainties, we add for each bin an additional
uncertainty of the same relative size as is given for the corresponding \RDDs{} measurement, which we assume to be uncorrelated between the different bins.
We expect this treatment to be conservative, given that we consider here the shape of the distributions and the systematic uncertainties typically show sizable correlations between the bins.

\begin{table}
\centering{
\caption{\it \small  Measured $q^2$ distributions for $B\to D^{(*)}\tau\nu$ events by Belle~\cite{Huschle:2015rga}. The fit
values with the normalization corresponding to the BaBar data~\cite{Lees:2013uzd} can be obtained by multiplying the given values by $1.4$ for $B\to D\tau\nu$ and $2.17$ for $B\to
D^*\tau\nu$.\label{tab:distBelle}}
\begin{tabular}{|c|c|c|c|c|}
\hline \rowcolor{RGray}
$q^2$ (GeV$^2$) & $B \to D \tau \nu$ & fit & $B \to D^* \tau \nu$  & fit \\
\hline 
$4.0-4.53$    & $24.0\pm16.3$   & $11\pm2$ 
                & $5.4\pm9.3$  & $\phantom{1}5.1 \pm 0.8$ 
                \\   \rowcolor{Gray}
$4.53-5.07$   & $27.8\pm15.2$   & $16\pm4$
                & $3.4\pm8.1$  & $\phantom{1}8.7\pm1.3$ 
                \\
$5.07-5.6$    & $22.0 \pm14.0$  & $20\pm5$
                & $-3.8\pm6.8$ & $12.1\pm1.8$ 
                \\ \rowcolor{Gray}
$5.6-6.13$    & $28.4 \pm 14.4$ & $24\pm6$
                & $12.1\pm8.4$ & $14.7\pm2.2$ 
                \\
$6.13-6.67$   & $16.2 \pm 14.8$ & $26\pm6$
                & $8.0\pm9.4$  & $16.8\pm2.5$ 
                \\ \rowcolor{Gray}
$6.67-7.2$    & $44.5 \pm15.5$  & $26\pm6$
                & $24.7\pm8.2$ & $18.3\pm2.7$ 
                \\
$7.2-7.73$    & $14.2 \pm 16.3$ & $27\pm6$
                & $2.7\pm7.8$  & $19.2\pm2.8$ 
                \\ \rowcolor{Gray}
$7.73-8.27$   & $-3.1 \pm 15.3$ & $26\pm6$
                & $28.7\pm9.2$ & $19.4\pm2.8$ 
                \\
$8.27-8.8$    & $16.1 \pm 15.2$ & $25\pm5$
                & $30.8\pm8.5$ & $18.9\pm2.8$ 
                \\ \rowcolor{Gray}
$8.8-9.33$    & $37.2\pm15.5$   & $23\pm5$
                & $24.9\pm7.6$ & $17.6\pm2.6$ 
                \\
$9.33-9.86$   & $19.3 \pm 15.2$ & $20\pm5$
                & $15.0\pm6.8$ & $15.4\pm2.4$ 
                \\ \rowcolor{Gray}
$9.86-10.4$   & $37.0\pm15.5$   & $17\pm4$
                & $14.8\pm5.1$ & $11.6\pm1.8$ 
                \\
$10.4-10.93$  & $-1.0 \pm 14.2$ & $13\pm3$
                & $16.3\pm5.1$ & $\phantom{1}3.6\pm0.6$ 
                \\ \rowcolor{Gray}
$10.93-11.47$ & $20.0\pm13.1$   & $\phantom{1}8\pm3$
                & --& --\\
$11.47-12.0$  & $3.4 \pm 10.9$  & $1.1\pm0.4$
                & --  & --\\
\hline
\end{tabular}
}\end{table}

\begin{table}
\centering{
\caption{\it \small Measured $q^2$ distributions for $B\to D^{(*)}\tau\nu$ events by BaBar~\cite{Lees:2013uzd}.\label{tab:distBaBar}}
\begin{tabular}{|c|c|c|}
\hline \rowcolor{RGray}
$q^2$ (GeV$^2$) & $B \to D \tau \nu$ & $B \to D^* \tau \nu$  \\
\hline
$4.0-4.5$    & $23.8 \pm 12.1$ &$0.6\pm7.1$  \\   \rowcolor{Gray}
 $4.5-5.0$  & $16.8\pm11.8$ & $23.6\pm9.5$  \\
$5.0-5.5$ & $27.9\pm10.5$ & $22.4\pm7.7$  \\ \rowcolor{Gray}
  $5.5-6.0$  & $45.1\pm13.1$ & $20.8\pm7.8$  \\
  $6.0-6.5$ & $46.9\pm13.3$ & $20.0\pm7.5$  \\ \rowcolor{Gray}
  $6.5-7.0$ & $39.7\pm13.6$ & $38.8\pm8.6$  \\
    $7.0-7.5$ & $31.7\pm12.4$ & $44.4\pm9.2$  \\ \rowcolor{Gray}
  $7.5-8.0$ & $47.4\pm14.9$ & $49.3\pm10.3$  \\
    $8.0-8.5$ & $33.7\pm14.0$ & $40.0\pm9.4$  \\ \rowcolor{Gray}
      $8.5-9.0$ & $17.7\pm13.2$ &   $37.3\pm9.5$  \\
          $9.0-9.5$ & $-0.7\pm13.1$ & $38.4\pm9.8$ \\ \rowcolor{Gray}
      $9.5-10.0$ & $6.9\pm14.3$ & $31.7\pm11.0$  \\
     $10.0-10.5$ & $35.4\pm16.0$ & $31.9\pm10.5$   \\ \rowcolor{Gray}
     $10.5-11.0$ & $2.8\pm12.1$ & $16.7\pm10.4$   \\
     $11.0-11.5$ & $1.7\pm11.3$ & --  \\ \rowcolor{Gray}
 $11.5-12.0$ & $6.5\pm8.9$ & --  \\
\hline
\end{tabular}
 }
\end{table}

The LEP experiments give an averaged constraint on $b\to X \tau\nu$~\cite{Olive:2016xmw},
\begin{equation}
\mathrm{Br}(b\to \tau\nu+{\rm anything})=(2.41\pm0.23)\%\,.
\end{equation}
This measurement is dominated by $b\to X_c\tau\nu$ because of $|V_{ub}|^2/|V_{cb}|^2\sim 1\%$. Correcting for the $b\to u$ contribution which is about $2\%$ due to the larger available phase space, we obtain
\begin{equation}
\mathrm{Br}(b\to X_c\tau\nu)=(2.35\pm0.23)\%\,.
\end{equation}
The LEP measurement corresponds to a known admixture of initial states for the weak decay~\cite{Abbaneo:2001bv}. The inclusive decay
rate does, however, not depend on this admixture to LO in $\Lambda_{\rm QCD}/m_b$. The corrections to this limit \emph{are}
hadron-specific and only partly known~\cite{Falk:1994gw,Grossman:1994ax}. It is again advantageous to consider the ratio $R(X_c)$,
defined in analogy to Eq.~\eqref{eq:RDdef} and cancelling again the $m_b^5 |V_{cb}|^2$ dependence. The scalar interactions in
Eq.~\eqref{eq:Lag} modify the inclusive decay width $\Gamma(b\to c \tau \bar\nu)$. Ignoring QCD corrections, we find
%
\beqn \label{eq::GammaLO}
\Gamma ( b\to c\tau\bar\nu) & = & \Gamma_{\!\!cb}\;\int_{x_{\tau}}^{(1-\sqrt{x_c})^2} dz\; \left( 1 -\frac{x_{\tau}}{z}\right) \lambda^{1/2}(1,x_c,z) \nonumber\\ &&
\hspace{-2.0cm}\times\;\Biggl\{ 2\,\left[ (1-x_c)^2 + z\, (1+x_c) - 2 z^2\right]
\nonumber\\ &&
\hspace{-1.6cm}+ 2\,\frac{x_{\tau}}{z}\,\left[ (1-x_c)^2 -2 z\, (1+x_c) + z^2\right] \nonumber\\ &&
\hspace{-1.6cm}- 2\, \frac{x_{\tau}^2}{z^2}\,\left[2\, (1-x_c)^2 -z\, (1+x_c) - z^2\right] \nonumber\\ &&
\hspace{-1.6cm}+ 6\,\mathrm{Re}(g_L^{cb\tau}) x_{\tau}^{1/2} x_c^{1/2}\left[ 1-x_c-x_{\tau} + z -\frac{x_{\tau}}{z} (1-x_c)\right] \nonumber\\ &&
\hspace{-1.6cm}+ 6\,\mathrm{Re}(g_R^{cb\tau})\, x_{\tau}^{1/2} \,\left[ 1-x_c+x_{\tau} - z -\frac{x_{\tau}}{z}\, (1-x_c)\right]
\nonumber\\ &&
\hspace{-1.6cm}+ 3\,\left( |g_L^{cb\tau}|^2 + |g_R^{cb\tau}|^2\right)\, (1+x_c-z)\, (z-x_{\tau}) \nonumber\\ &&
\hspace{-1.6cm}+ 12\,\mathrm{Re}(g_L^{cb\tau \,*} g_R^{cb\tau})\, x_c^{1/2}\, (z-x_{\tau}) \Biggr\}\,,
\eeqn
%
where $\Gamma_{\!\!cb}=\frac{G_F^2 m_b^5}{192\pi^3}\, |V_{cb}|^2$, $x_{\tau} =m_{\tau}^2/m_b^2$, $x_c =m_c^2/m_b^2$, $z =q^2/m_b^2$ and $\lambda(x,y,z)= x^2+y^2+z^2-2(xy+yz+xz)$. Here $q^2 = (p_{\tau} +
p_{\nu})^2 = (p_b-p_c)^2$ is the invariant mass squared of the lepton pair. These results confirm known SM expressions at this
order~\cite{Koyrakh:1993pq,*Balk:1993sz,*Bigi:1993fe,*Blok:1993va,*Manohar:1993qn,Falk:1994gw} and generalize the results of
Refs.~\cite{Grossman:1994ax,Grossman:1995yp,Czarnecki:1992zm} for 2HDMs with NFC, which we reproduce in the corresponding limit. We can rewrite the differential decay rate as follows:
\begin{eqnarray}
\frac{d\Gamma ( b\to c\tau\bar\nu)}{dz} & = & \Gamma_W+\frac{3\,\Gamma_{\!\!cb}}{4}\sqrt{\lambda(1,z,x_c)}\times\\
&&\left[(1-z+x_c)\,f_1(z)+2\sqrt{x_c}\,f_2(z)\right]\,,\nonumber
\end{eqnarray}
where the functions $f_{1,2}(z)$ are given as
\begin{align}
f_i(z)=\sum_{I,J=G,H}h_i^{IJ}L_{IJ}(z)\,,
\end{align}
with $h_1^{IJ}=\left(a_I a_J^*+b_I b_J^*\right)$, $h_2^{IJ}=\left(a_I a_J^*-b_I b_J^*\right)$, $h_3^{IJ}={\rm Re}(a_I b_J^*)$ and
$L_{IJ}(z) = z (1-x_\tau/z)^2\left(a_I^\ell a_J^{\ell *}+b_I^\ell b_J^{\ell *}\right)$. This formulation shows explicitly that the
decay rate is a sum of two incoherent terms, the first of which corresponds to a transverse $W$ exchange, while the second stems
from both the charged-scalar and longitudinal $W$ exchanges. Splitting the phase space for the latter as the product of that of $b\to c
W^*$ and that of $W^*\to \tau\nu$ together with calculating it in Landau gauge allows to obtain the result for the charged-scalar
contribution from the known calculations for $t\to b W$ and $t\to b H$~\cite{Jezabek:1988iv,*Jezabek:1988ja,Czarnecki:1992zm} by
identifying the changed couplings and propagators~\cite{Czarnecki:1994bn,Grossman:1995yp}.

The $\mathcal O(\alpha_s)$ corrections are given as
\begin{eqnarray}
\left.\frac{d\Gamma ( b\to c\tau\bar\nu)}{dz}\right|_{\alpha_s} & = &
\Gamma_W^{\alpha_s}+\frac{2\alpha_s}{\pi}\Gamma_{\!\!cb}\times\left[ G_+(z)f_1(z)\right.\\
&&\left.+\sqrt{x_c}G_-(z)f_2(z)+G_0(z) f_3(z)\right]\,,\nonumber
\end{eqnarray}
where the functions $G_i(z)$ can be found in Ref.~\cite{Czarnecki:1992zm} and the transverse-$W$ contribution $\Gamma_W^{\alpha_s}$ in
Ref.~\cite{Czarnecki:1994bn}. These expressions generalize the existing ones in 2HDMs with NFC~\cite{Kalinowski:1990ba,Grossman:1994ax,Grossman:1995yp}.

The functions $h_i^{IJ}$ determine the relative strengths of charged-scalar and (longitudinal) $W$ exchanges, as well as their interference; only their overall
coefficients can change at higher orders. At LO only two combinations of couplings appear, despite the presence of four combinations of
$g_{L,R}$ in Eq.~\eqref{eq::GammaLO}. At NLO, a third combination enters, to be compared with five independent combinations when written as in Eq.~\eqref{eq::GammaLO}.

The products of the charged-scalar couplings $(a,b)_H^{(\ell)}$ correspond to the couplings in the effective Lagrangian as follows:
\begin{equation}
a_H (a_H^\ell)^{\ast} = -(g_L^{cb\tau}+g_R^{cb\tau})\,, b_H (b_H^\ell)^{\ast} = -(g_L^{cb\tau}-g_R^{cb\tau})\,,
\end{equation}
and the Goldstone couplings are given as
\begin{equation}
a_G (a_G^\ell)^{\ast} = -\frac{(1-\sqrt{x_c})\sqrt{x_\tau}}{z}\,, b_G (b_G^\ell)^{\ast} = \frac{(1+\sqrt{x_c})\sqrt{x_\tau}}{z}\,.
\end{equation}
Additionally the relations $a_H^\ell=-b_H^\ell$ and $a_G^\ell=-b_G^\ell$ hold due to the neglect of neutrino masses.

Numerically, it turns out that the $\mathcal O(\alpha_s)$ corrections cancel largely in the SM part of the ratio $R(X_c)$, yielding a rather small correction of below $3\%$; therefore we do not include higher-order corrections which are known only in the SM, show similar cancellations and are correspondingly
smaller~\cite{Biswas:2009rb}. The shift for the NP couplings is however larger, making the different coefficients of the $f_i(z)$ receive a significant reduction at NLO of about $30\%$.
Using the $1S$ mass scheme~\cite{Hoang:1998ng,Hoang:1998hm}, we obtain schematically
\begin{eqnarray}\label{eq::RXc}
R(X_c)=0.231&&\left[c_{\rm SM}+(0.183\, c_1-0.050\, c_3)|g^{cb\tau}_L|^2\right.\\
&&\hspace{-0.4cm}+0.183\,c_1|g^{cb\tau}_R|^2+0.278\, c_2\,{\rm{Re}}(g^{cb\tau}_Lg_R^{cb\tau*})\nonumber\\
&&\hspace{-0.4cm}+(0.296\,c_2-0.117\,c_1+0.030\,c_3)\,{\rm{Re}}(g^{cb\tau}_L)\nonumber\\
&&\hspace{-0.4cm}\left.+(0.404\,c_1-0.086\,c_2+0.109\,c_3)\,{\rm{Re}}(g^{cb\tau}_R)\right]\,,\nonumber
\end{eqnarray}
to be compared with the LO expression
\begin{eqnarray}
\hspace{-0.4cm}
R(X_c)=0.224&&\left[c_{\rm SM}+0.250\,c_1\left(|g^{cb\tau}_L|^2+|g^{cb\tau}_R|^2\right)\right.\\
&&+0.396\, c_2\,{\rm{Re}}(g^{cb\tau}_Lg_R^{cb\tau*})\nonumber\\
&&+(0.421\,c_2-0.152\,c_1)\,{\rm{Re}}(g^{cb\tau}_L)\nonumber\\
&&\left.+(0.548\,c_1-0.117\,c_2)\,{\rm{Re}}(g^{cb\tau}_R)\right]\,.\nonumber
\end{eqnarray}
The factors $c_i$ are introduced as $f_i(z)\to c_i\,f_i(z)$, in order to track the corresponding correlations between the different NP contributions;
they are varied around their central values $c_i=1$ in the numerical analysis in order to account for the presence of higher-order contributions. Note that apart from the sizable numerical shift in the coefficients, there is also a qualitative difference between the expressions at LO and NLO: at LO, the coefficients of $|g^{cb\tau}_{L,R}|^2$ are equal, which leads to the absence of interference terms between $\delta_{cb}^{\tau}$ and $\Delta_{cb}^{\tau}$,
which allows to write $R(X_c)$ as a sum of positive definite terms, leaving no possibility for cancellations. This is not true at NLO, and therefore
strictly speaking there is no constraint in the individual $\delta_{cb}^{\tau}$ and $\Delta_{cb}^{\tau}$ planes without restricting the combination not shown.

Finally, we also include the SM power corrections of order $\Lambda_{\rm QCD}^2/m_b^2$. They have been calculated for the $B$-meson decay in
Ref.~\cite{Falk:1994gw} and amount to $\sim 4\%$ of the NLO value for $R(X_c)$. $\mathrm{SU(3)}$ symmetry predicts them to be equal for
$B_{u,d,s}$ mesons, which contribute $\sim 90\%$  in the LEP measurement. The (unknown) shift to the power corrections for $\Lambda_b$
decays as well as the corrections to the $\mathrm{SU(3)}$ assumption  are included as an uncertainty of the leading term in
Eq.~\eqref{eq::RXc}.
This reduces the SM value from $0.231$ to $0.222$, which is in agreement with the result in Ref.~\cite{Ligeti:2014kia}.


\section{Fit details}  \label{fitdetl}

In Table~\ref{tab::chi2} we collect details of the fit performed for the different benchmark scenarios.


\begin{table}
\centering
\caption{Minimal-$\chi^2$ values obtained in the considered scenarios, given for different sets of observables, together with the corresponding central values for the NP parameters. Note that the central values are only given for
illustration. \label{tab::chi2}} \vspace{0.2cm}\doublerulesep 0.8pt \tabcolsep 0.05in
\begin{tabular}{l c c c c}\hline\hline  \rowcolor{white}
Scenario  & $\chi^2_{\rm min}$  & \# obs.  & \# pars. & central values ($\delta_{cb}^\tau$, $\Delta_{cb}^\tau$)\\  \hline
\multicolumn{5}{c}{\RDDs{} only}\\ \hline
SM  & 23.1  & 2  & 0  & ---\\ \rowcolor{Gray}
S1  & 0  & 2  & 4 &  $(0.2+0.7i,10.0-6.3i)$\\
S1 real  & 0  & 2  & 2  & $(0.4,-3.6)$\\ \rowcolor{Gray}
$g_L^{cb\tau}$  & 0  & 2  & 2  & $g_L^{cb\tau}=-1.3-0.6i$\\
$g_R^{cb\tau}$  & 9.1  & 2  & 2  & $g_R^{cb\tau} = 0.3+0.i$\\ \rowcolor{Gray}
$g_{V_L}$  & 0.2  & 2  & 1  & $|g_{V_L}|=1.12$\\ \hline
\multicolumn{5}{c}{\RDDs{}, $d\Gamma/dq^2$, $\Gamma_{B_c}$}\\
\hline
SM  & 65.9  & 61  & 4  & ---\\ \rowcolor{Gray}
S1  & 49.2  & 61  & 8  & $(0.4+0.i,-2.4+0.i)$\\
S1 real  & 49.2  & 61  & 6  & $(0.4,-2.4)$\\ \rowcolor{Gray}
$g_L^{cb\tau}$  & 55.4  & 61  & 6  & $g_L^{cb\tau}=-0.4+0.8i$\\
$g_R^{cb\tau}$  & 55.4  & 61  & 6  & $g_R^{cb\tau} = 0.3+0.i$\\ \rowcolor{Gray}
$g_{V_L}$  & 42.4  & 61  & 5  & $|g_{V_L}|=1.12$\\  \hline
\multicolumn{5}{c}{\RDDs{}, $d\Gamma/dq^2$, $\Gamma_{B_c}$, $R(X_c)$}\\
\hline
SM  & 65.9  & 62  & 4  & ---\\ \rowcolor{Gray}
S1  & 50.4  & 62  & 8  & $(0.3+0.i,-2.4+0.i)$\\
S1 real  & 50.4  & 62  & 6  & $(0.3,-2.4)$\\ \rowcolor{Gray}
$g_L^{cb\tau}$  & 55.4  & 62  & 6  & $g_L^{cb\tau}=-0.4-0.8i$\\
$g_R^{cb\tau}$  & 56.1  & 62  & 6  & $g_R^{cb\tau} = 0.2+0.i$\\ \rowcolor{Gray}
$g_{V_L}$  & 46.7  & 62  & 5  & $|g_{V_L}|=1.10$
\\\hline\hline
\end{tabular}
\end{table}

\bibliography{sl15}

\end{document}